\documentclass[longauth]{aa}  

\usepackage{graphicx}

\usepackage{txfonts}

\usepackage[colorlinks = true,
           linkcolor = blue,
            urlcolor  = blue,
            citecolor = blue,
            anchorcolor = blue]{hyperref}

\newcommand{\oiii}{\ifmmode \text{[O~{\sc iii}]} \else[O~{\sc iii}]\fi}

\newcommand{\ha}{\ifmmode \text{H}\alpha \else H$\alpha$\fi}
\newcommand{\hb}{\ifmmode \text{H}\beta \else H$\beta$\fi}
\newcommand{\oiiihb}{\text{\oiii+\hb}}
\newcommand*\standardrel{<}
\newcommand*\standardbin{+}
\makeatletter
\newcommand*\tabularrel[1]{%
  \mathrel{\mathpalette{\@tabularsym\standardrel}{#1}}%
}
\newcommand*\tabularbin[1]{%
  \mathbin{\mathpalette{\@tabularsym\standardbin}{#1}}%
}
\newcommand*\@tabularsym[3]{%
  \setbox\z@\hbox{$#2#1\m@th$}%
  \hbox to\wd\z@{\hss$#2#3\m@th$\hss}%
}

\usepackage{xcolor}
\usepackage{natbib}

\newcommand{\kms}{\rm{km\ s}^{-1}} 
\begin{document} 
   \title{JWST COSMOS-3D: Spectroscopic census and luminosity function of \oiii\  emitters at $6.75\!<\!z\!<\!9.05$ in COSMOS}
   \titlerunning{JWST COSMOS-3D: \oiii\ Census and luminosity function}
   \author{Romain A. Meyer
          \inst{1}
          \and
          Feige Wang \inst{2}
          \and
          Koki Kakiichi \inst{3}
          \and 
          Gabe Brammer \inst{3} 
          \and
          Jackie Champagne \inst{4} 
          \and 
          Katharina Jurk \inst{5} 
          \and
          Zihao Li \inst{3}
          \and 
          Zijian Li  \inst{6}
          \and  
          Marat Musin  \inst{6}
          \and
          Sindhu Satyavolu  \inst{7}
          \and
          Jan-Torge Schindler  \inst{5}
          \and
          Marko Shuntov \inst{3,8}
          \and
          Yi Xu \inst{3,9}
          \and
          Siwei Zou  \inst{6,10}  
          \and  
          Fuyan Bian \inst{11,6}
          \and
          Caitlin Casey \inst{12,3}
          \and 
          Eiichi Egami \inst{4}
          \and
          Xiaohui Fan \inst{4}
          \and
          Danyang Jiang \inst{13,14}
          \and 
          Nicolas Laporte \inst{15}
          \and
          Weizhe Liu \inst{4}
          \and
          Pascal A. Oesch \inst{1,3}
          \and
          Lidia Tasca \inst{15}
          \and 
          Jinyi Yang \inst{2}
          \and
          Zijian Zhang \inst{13,14}
          \and 
          Hollis Akins \inst{16}
          \and
          Zheng Cai \inst{17}
          \and
          Dave A. Coulter \inst{18,19}
          \and
          Jiamu Huang \inst{12}
          \and
          Mingyu Li \inst{17}
          \and
          Weizhe Liu \inst{4}
          \and
          Yongming Liang \inst{20,21}
          \and
          Bingcheng Jin \inst{2}
          \and
          Xiangyu Jin \inst{2} 
          \and
          Jeyhan Kartaltepe \inst{22}
          \and
          Anton M. Koekemoer \inst{18}
          \and
          Jasleen Matharu \inst{23}
         \and
          Maria Pudoka \inst{4}
          \and
          Wei-Leong Tee \inst{24}
          \and 
          Callum Witten \inst{1}
          \and
          Haowen Zhang \inst{25}
          \and
          Yongda Zhu \inst{4}
          }

   \institute{Department of Astronomy, University of Geneva, Chemin Pegasi 51, 1290 Versoix, Switzerland\\
              \email{romain.meyer@unige.ch}
         \and
             Department of Astronomy, University of Michigan, 1085 S. University Ave., Ann Arbor, MI 48109, USA
        \and 
            Cosmic Dawn Center (DAWN), Niels Bohr Institute, University of Copenhagen, Jagtvej 128, DK-2200 Copenhagen N, Denmark
        \and
        Steward Observatory, University of Arizona, 933 N. Cherry Ave, Tucson, AZ 85721, USA
        \and 
        Hamburger Sternwarte, Universit\"at Hamburg, Gojenbergsweg 112, D-21029 Hamburg, Germany
        \and 
        Chinese Academy of Sciences South America Center for Astronomy, National Astronomical Observatories, CAS, Beijing 100101, China
        \and
        Institut de Física d’Altes Energies (IFAE), The Barcelona Institute of Science and Technology, Edifici Cn, Campus UAB, 08193, Bellaterra (Barcelona), Spain
        \and 
        University of Geneva, 24 rue du Général-Dufour, 1211 Genève 4, Switzerland
        \and
        National Astronomical Observatory of Japan, 2-21-1 Osawa, Mitaka, Tokyo 181-8588, Japan 
        \and
        Departamento de Astronom\'ia, Universidad de Chile, Casilla 36-D, Santiago, Chile
        \and
        European Southern Observatory, Alonso de C\'ordova 3107, Casilla 19001, Vitacura, Santiago 19, Chile
        \and
        Department of Physics, University of California, Santa Barbara, Santa Barbara, CA 93106, USA
        \and 
        Department of Astronomy, School of Physics, Peking University, Beijing 100871, China
        \and
        Kavli Institute for Astronomy and Astrophysics, Peking University, Beijing 100871, China
        \and 
        Aix Marseille Universit\'e, CNRS, CNES, LAM (Laboratoire d’Astrophysique de Marseille), UMR 7326, 13388 Marseille, France
        \and
        Department of Astronomy, The University of Texas at Austin, 2515 Speedway Blvd Stop C1400, Austin, TX 78712, USA
        \and 
        Department of Astronomy, Tsinghua University, Beijng, China 
        \and
        Space Telescope Science Institute, 3700 San Martin Drive, Baltimore, MD 21218, USA
        \and
        Physics and Astronomy Department, Johns Hopkins University, Baltimore, MD 21218, USA
        \and
        Institute for Cosmic Ray Research, The University of Tokyo, 5-1-5 Kashiwanoha, Kashiwa, Chiba 277-8582, Japan
        \and
        National Astronomical Observatory of Japan, 2-21-1 Osawa, Mitaka, Tokyo 181-8588, Japan
        \and 
        Laboratory for Multiwavelength Astrophysics, School of Physics and Astronomy, Rochester Institute of Technology, 84 Lomb Memorial Drive, Rochester, NY 14623, USA
        \and
        Max-Planck-Institut f\"ur Astronomie, K\"onigstuhl 17, D-69117 Heidelberg, Germany
        \and
        Department of Astronomy and Astrophysics, The Pennsylvania State University, 525 Davey Lab, University Park, PA 16802, USA
        \and
        Canadian Institute for Theoretical Astrophysics, 60 St George St, Toronto, ON M5S 3H8, Canada
    }

   \date{Received October 13, 2025; accepted -- }

 \abstract{We present a catalogue of spectroscopically selected \oiiihb\ emitters at $6.75\!<\!z\!<\!9.05$ and the resulting \oiii\ $5008$ \AA \ luminosity function (LF) in the COSMOS field. We leveraged the 0.33 deg$^{2}$ covered  by COSMOS-3D using NIRCam/WFSS F444W to perform the largest spectroscopic search for \oiii\ emitters at $6.75\!<\!z\!<\!9.05$. We present our catalogue of $249$ \oiii\ emitters and their associated completeness function. The inferred constraints on the \oiii\ LF enabled us to characterise the knee of the \oiii\ $5008\ \AA$ LF, resulting in improved \oiii\ $5008\ \AA$ LF constraints at $z\sim 7, 8$. Notably, we find evidence for a decline of the \oiii\ luminosity density between $z\sim6-8$, which could be expected if the metallicity of \oiii\ emitters, as well as the cosmic star-formation rate density, is declining at these redshifts. We find that theoretical models that reproduce the $z\sim7,8$ \oiii\ $5008\ \AA$ LF do not reproduce well the \oiii\ equivalent width distribution, pointing to potential challenges in the modelling of \oiii\ and other nebular lines in the early Universe. Finally, we provide the first constraints on the cosmic variance of \oiii\ emitters, estimating at $15\%$ the fractional uncertainty for the z$\sim 7,8$ \oiii\ $5008\ \AA$ LF in the 0.33 deg$^2$ field. This estimate is in good agreement with that inferred from clustering, and it shows that the \oiii\ $5008\ \AA$ LF derived from smaller extragalactic legacy fields is strongly affected by cosmic variance. Our results highlight the fundamental role that wide-area JWST slitless surveys play to map the galaxy large-scale structure down into the reionisation era, serving as a springboard
 for a variety of science cases. }
   \maketitle
   \nolinenumbers

\section{Introduction}
Understanding the formation of the first galaxies and the emergence of the first large-scale structures in the Universe is a fundamental goal of modern astronomy and astrophysics. The \textit{James Webb Space Telescope} (JWST) has transformed our ability to study the first billion years of the Universe with its unprecedented sensitivity and access to the near-infrared regime. Large imaging campaigns have now established a first census of photometrically selected galaxies up to $z\sim 14.5$ \citep{Eisenstein2023,Donnan2023,Donnan2024, Robertson2024, Finkelstein2024} and even candidates at $z\gtrsim 15 -20$ \citep{Perez-Gonzalez2025,Kokorev2025}. Compared to the \textit{Hubble Space Telescope} (HST) era, spectroscopic confirmations at $z>10$ and up to $z\sim15$ are now routine \citep{Bunker2023_GNz11,CurtisLake2023,Carniani2024,Naidu2025}. Rest-frame ultra-violet (UV) and optical spectroscopy has opened a new window into the properties  of early luminous galaxies, revealing hard radiation fields, low metallicity, active galactic nuclei (AGN) activity, and bursty star-formation histories \citep[e.g.][]{Tang2025,Ellis2025,RobertsBorsani2026}.

High-redshift cosmography (i.e. the science of mapping the Universe) has also been transformed by JWST. In particular, NIRCam \citep{2023PASP..135b8001R} Wide Field Slitless Spectroscopy (WFSS) programmes (e.g. FRESCO \citealt{Oesch2023}, CONGRESS (Egami, Sun in prep.), ALT \citealt{Naidu2024}, NEXUS \citealt{Shen2024_NEXUS}, EIGER \citealt{Kashino2023}, ASPIRE \citealt{Wang2023}) have provided increasingly spectroscopically complete samples of galaxies in key extragalactic deep fields, lensing fields and quasar fields. In turn, WFSS campaigns have enabled the accurate censuses of line emitters at $z>5$ \citep{Meyer2024,CoveloPaz2025}, the discovery of little red dots \citep[LRD, e.g.][]{Matthee2024_LRD,Lin2024}, unveiled particular overdensities and proto-cluster candidates \citep{Sun2024,Helton2024b, Herard-Demanche2025,Witten2025, Champagne2025}, and provided clustering measurements with spectroscopic redshifts out to $z\sim8$ for the first time \citep{Shuntov2025_clustering}. 

Although these WFSS programmes provide a transformative view of the statistics of early galaxies and their environments, their field of view is still limited (in the best cases) to scales $\lesssim 10\ \rm{arcmin}^{2}$, or $\lesssim 0.4\ \rm{cMpc}$ at $6\!<\!z\!<\!9$. However, the true promise of high-redshift cosmography is to map the emergence of the cosmic web and its connection with the topology of reionisation, which requires mapping galaxies at least over tens of megaparsecs \citep[e.g.][]{Lee2014, Horowitz2022, Kakiichi2023}. COSMOS-3D is a 268h JWST programme mapping a significant fraction ($0.33$ deg$^{2}$) of the legacy Cosmic Evolution Survey \citep[COSMOS,][]{Scoville2007,Capak2007} field with NIRCam WFSS F444W (Kakiichi in prep.). COSMOS-3D builds upon the legacy of imaging in the COSMOS field, in particular the $0.54$ deg$^2$ JWST/NIRCam imaging in F115W, F150W, F277W $\&$ F444W of COSMOS-Web \citep[][]{Casey2023} and the initial COSMOS survey \citep{Scoville2007,Capak2007}. With $\sim10\times$ the area of the First Reionization Epoch Spectroscopically Complete Observations (FRESCO) survey, but only $2\times$ shallower, and a total survey volume of $5\times10^7 \ \rm{cMpc}^{3}$ over $6.75\!<\!z\!<\!9.0$, COSMOS-3D provides a significant step in the area and scales covered, as well as a large increase in the expected number of spectroscopic redshifts obtained. COSMOS-3D is therefore poised to transform our understanding of the first structures (e.g. Champagne et al., in prep.), the statistics of the first galaxies, and reveal rare objects in the early Universe \citep{Lin2025_LRD, Tanaka2025}.

In this paper, we present the result of a search for \oiiihb\ emitters at $6.75\!<\!z\!<\!9.0$ in the first $90\%$ of the COSMOS-3D data. The search was optimised to provide a well-characterised, end-to-end selection function for the resulting sample, which is crucial for statistical analyses such as the luminosity function (LF) and clustering measurements. Additionally, we used the same analysis method as in \citet{Meyer2024}, thus facilitating comparisons and common analysis with the GOODS fields surveyed by FRESCO. In Section \ref{sec:method}, we summarise our data reduction and selection method, before discussing the general properties of the sample and our \oiii\ constraints in Section \ref{sec:results}. We leverage the $10\times$ larger area of COSMOS-3D compared to previous surveys (e.g. FRESCO) to constrain the impact of cosmic variance in Section \ref{sec:discussion} before concluding in Section \ref{sec:conclusion}.

Throughout this paper, magnitudes are given in the AB system \citep[][]{Oke1974}{}{}, and we assume a concordance cosmology with $H_0 = 70\ \kms \rm{Mpc}^{-1},\ \Omega_m = 0.3, \Omega_\Lambda=0.7$. Errors are given for the $16-84$ percentiles or the standard deviation if single-valued.

\section{Method}
\label{sec:method}
This work closely follows the method of \citet{Meyer2024} using similar NIRCam  WFSS F444W observation in the GOODS fields from the FRESCO survey \citep[][]{Oesch2023}. We summarise the relevant aspects of the reduction process and the main differences with respect to FRESCO as follows.
\subsection{COSMOS-3D NIRCam/WFSS data and reduction}
COSMOS-3D is a JWST Cycle 3 268h Large Programme (\# 5893, PIs: Kakiichi, Egami, Fan, Lyu,
Wang, and Yang) observing $0.33\ \rm{deg}^2$ in the COSMOS field \citep{Casey2023} with NIRCam WFSS in the F444W and NIRCam imaging in F115W, F200W, and F356W. Coordinated MIRI imaging parallels cover a final area of $484\ \rm{arcmin}^2$ in F1000W and F2100W. Further details of the COSMOS-3D programme will be presented in Kakiichi et al. (in prep.).

We reduced the COSMOS-3D NIRCam/WFSS data using \texttt{grizli}\footnote{\url{https://grizli.readthedocs.io/}} version 1.12.11 and \texttt{jwst} pipeline version 1.18, and CRDS context pmap 1293. \texttt{grizli} utilises from the \textit{uncal} files downloaded from \emph{MAST} and pre-processes the rate files by applying a modified step 1 of the \emph{jwst} pipeline using a custom snowball correction. We then produce WFSS images and spectral extractions using \texttt{grizli}'s framework \citep{Brammer2019, Brammer2021}, where we make use of the V9 spectral tracing and grism dispersion models for NIRCam WFSS \footnote{\url{https://github.com/npirzkal/GRISM_NIRCAM}}. In the grizli pipeline we applied median filtering along the dispersion direction to subtract the continuum of all sources in the field in a two-step manner. A median filter with a $71$ pixels wide kernel with a central gap of $20$ pixels is applied first. Pixels at signal-to-noise ratio S/N$>=3$ after median filtering are flagged and the resulting median filtered data discarded. The median filter is then applied again on the original data but with the S/N$>=3$ pixels masked in order to improve continuum oversubtraction around strong emission lines. After continuum subtraction we reached a median rms noise of $\sigma_{rms} = 5.0\times 10^{-18}\ \rm{erg\ s}^{-1}\rm{cm}^{-2}$ at $\lambda = 4\mu\rm{m}$ for a line width of $300\ \rm{km\ s}^{-1}$ (see further Appendix \ref{app:rms} for the varying sensitivity and coverage of the final WFSS mosaic). For each source, we used the F277W+F444W detection image made using the COSMOS-Web imaging data \citep{Shuntov2025_DR} to perform optimal extraction of the spectrum \citep{Horne1986}. 

\begin{figure}
    \centering
    \includegraphics[width=\linewidth]{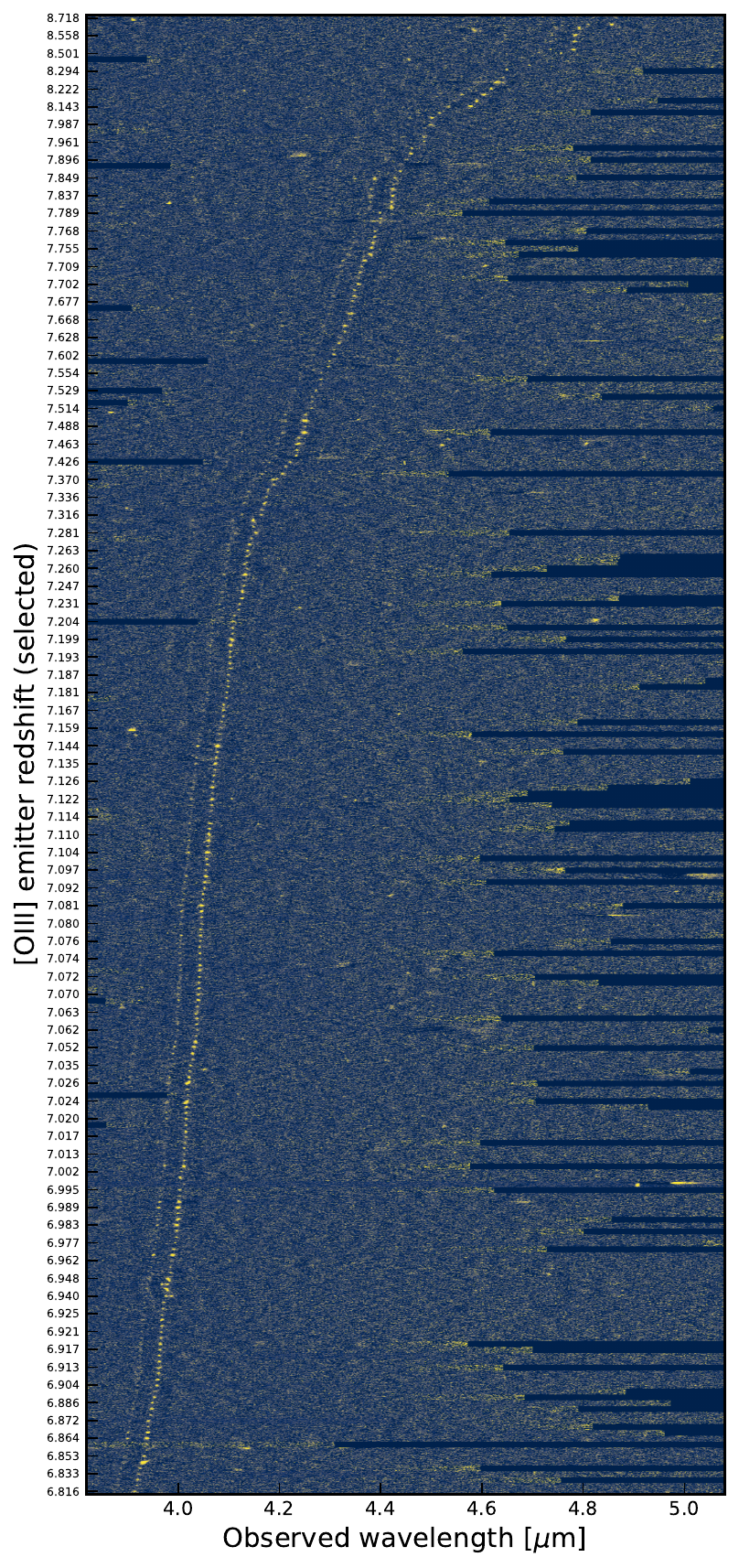}
    \caption{2D S/N spectra of all $249$ individual \oiii\ emitters reported in this work. The emitters are ordered by redshift (only a subsample of redshifts are shown on the y-axis for readability).}
    \label{fig:fig1_allspectra}
\end{figure}
\subsection{Selection of \oiii\ emitters candidates}
We pre-selected candidates from the COSMOS-Web v1 photometric catalogue \citep{Shuntov2025_DR} by removing any object that does not satisfy a strict Lyman-$\alpha$ break or is detected in the visible bands
\begin{align}
    & (\rm{S/N(F606W)} < 2\  \&\  \rm{S/N(F814W)} < 2 \ )  \\ &\rm{OR} \nonumber \\  & (\rm{S/N(F606W)} < 3\   \&\  \rm{S/N(F814W)} < 3 \nonumber \\ &\&\ \rm{F814W/F160W} < 0.2\  \&\ \rm{S/N(F160W)}>5) \text{\ \ \ .}
\end{align}
For parts of the field where F606W/F160W imaging is not available, we only requested a non-detection in F814W, where in all cases we made use of HST imaging data primarily from COSMOS \citep{Koekemoer2007} and CANDELS \citep{Grogin2011, Koekemoer2011}.
We further required a detection in the F444W band designed to remove objects too faint to be detectable with a single line and no continuum at the COSMOS-3D grism depth
\begin{align}
    & \rm{S/N(F444W)}>3\  \& \ m(\rm{F444W}) < 28 \text{\, \, \, ,}
\end{align}
where we used the aperture ($r=0.5^{"}$) magnitude from \citet{Shuntov2025_DR}.
This results in $92409$ candidates in the COSMOS-Web footprint. We then optimally extracted the spectrum for each object at least partially covered by COSMOS-3D. The 1D spectra are then analysed with a Gaussian-matched filter with full width half maximum (FWHM) $=50,100,200\ \rm{km\ s}^{-1}$ to extract significant emission line candidates. We then retained objects for which a) a candidate line doublet has a velocity separation matching that of \oiii\ $\lambda\lambda 5008, 4960$ \AA\ at $6.75\!<\!z\!<\!9.0$, with a tolerance $\Delta v < 100\ \kms$, b) the strongest line of the identified doublet is detected at S/N$>4$ and c) the doublet ratio of the gaussian-matched filtered lines is $1<  \oiii\ 5008 /   \oiii\ 4960 < 10$. Importantly, photometric redshift estimates are not used in the selection or during the subsequent visual inspection stage.

This yields a total of $2092$ objects of which $559$ are immediately removed as they are clearly contaminated by nearby stars, CO bandheads or their candidate doublet is clearly noise at the edge of the WFSS image. This leaves $1533$ candidates that were inspected visually with \texttt{specvizitor} \footnote{\url{https://ivkram.github.io/specvizitor/}}.
The visual inspection was carried out by a team of $10$ inspectors such that each object is inspected at least $3$ times and up to $5$ times. The candidates were distributed such that the sample inspected by each person is different and covers randomly the entire COSMOS-3D footprint in order to minimise spatially varying selection effects. Additionally, each inspector was instructed to start the visual inspection in a different part of the field in order to homogenise differential selection effects between the first and last object inspected. As in \citet{Meyer2024}, the objects are graded with five quality flags: $q=-1$ if the candidate line(s) is due to contamination by another object or continuum subtraction residuals, $q=0$ if no convincing line is observed, $q=1$ if only one convincing line, but no doublet, is observed, $q=2$ if a doublet at the right separation is observed, signalling a likely \oiii\ emitter, $q=3$ for a `definitive' \oiii\ emitter signalled either by a very high S/N detection of the doublet, a doublet and a H$\beta$ line detection, or a peculiar morphology common to the direct F444W imaging and the reconstructed \oiii\ line map making the identification certain. The catalogue is then screened for duplicates by comparing the x-y pixel position of the line emission in the grism data and the sky position of each detection. We also manually improve the median filtering by applying a custom mask to the \oiiihb\  lines, improving over the grizli pipeline default optimising for single line emission. Finally, we re-measured the \oiiihb\  lines fluxes and redshift by fitting three Gaussians with tied wavelengths and width (but leaving the three amplitudes free) using \emph{emcee} \citep{Foreman-Mackey2013}.

The final catalogue used in this work comprises 249 \oiiihb\ emitters at $6.75\!<\!z\!<\!9.0$ for which the median quality flag is $\geq 1.5$ suited for the statistical analysis present in this work. We note that for follow-up observations, a quality flag $q\geq$ 2 should be preferred if a near-perfect purity is required. We show the 2D spectra of all the sources in Fig. \ref{fig:fig1_allspectra} ordered by redshift. The full catalogue, 1D and 2D plots can be found in Appendix \ref{app:full_cat}.

\subsection{Completeness function}
We assessed the completeness of our search by injecting $320$ mock emitters in the median-subtracted WFSS data and recovering them through the same pipeline (guassian-matched filter and visual inspection). The mock emitters were inserted in visually vetted objects across the COSMOS-3D footprint passing our photometric quality cuts but rejected by the gaussian-matched (GM) filter such that their morphology resemble closely that of real emitters. The $320$ emitters span a grid of redshift ($6.75\leq z\leq9.05$ , with $\Delta z=0.15$), \oiii\ $5008\ \AA$ fluxes ($-18.3 \leq \log_{10} f_{\oiii\ 5008} \leq 16.4$, in $\Delta \log_{10}f = 0.1$), and \oiii\ / H$\beta$ ratio ($-1 \leq \log_{10} \oiii/\rm{H}\beta \leq 1.3$, in $\Delta = 0.1$ increments). The mock emitters were then mixed with the real galaxy candidates and inspected in a blind-trial fashion by the visual inspection team. We note that the above parameters include objects with H$\beta$/\oiii\ ratios up to $10$, i.e. inclusive of very metal-poor candidates. The presence strong H$\beta$ lines improves the recovery fraction of faint \oiii\ emitters even if the secondary line ($4960$ \AA) is not detected. Since no such objects are detected in the final sample, we only computed the completeness using mock sources with H$\beta$/\oiii\ $<3$ (corresponding to a total of 225 mock emitters) as this would otherwise bias our inferred completeness function high.

\begin{figure}
    \centering
    \includegraphics[width=\linewidth]{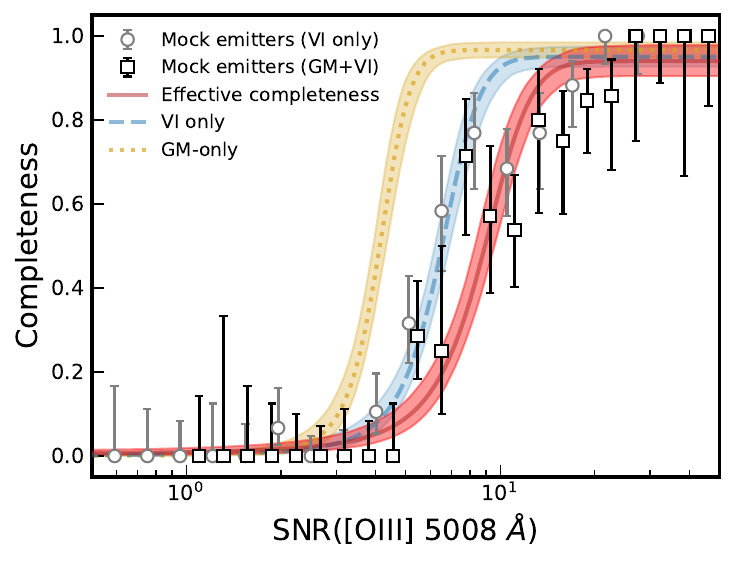}
    \caption{Completeness functions for the \oiiihb\ emitter search as a function of the measured \oiii\ 5008 $\AA$ S/N. The black square datapoints and red curve denote the effective completeness including the initial Gaussian-matched filtering (GM) and the subsequent visual inspection (VI). The best-fit completeness function of each separate step is shown with yellow and blue lines, as well as the binned data values for the visual inspection (grey dots). Note that the visual inspection is specific to the quality threshold chosen (here $q\geq1.5$). }    
    \label{fig:completeness_function}
\end{figure}

By applying the selection criteria detailed in the previous section, we determined the end-to-end completeness of our spectroscopic search, which we show in Fig. \ref{fig:completeness_function}. The completeness is well-characterised by a sigmoid function 
\begin{equation}
    C(\rm{S/N}) = \frac{C_{max}}{1+\exp{(-a * (\rm{S/N} - \rm{S/N}_0))}}\label{eq:c_spec} \text{\ \ \ \ ,}
\end{equation}
with the best-fit parameters $C_{\rm{max}} = 0.94\pm0.04$, $a=0.6\pm0.12$, $\rm{S/N}_0=8.72\pm0.66$. We note that a $50\%$ completeness is achieved at S/N(\oiii)$= 8.7$, corresponding to a detection of the fainter member  at S/N$\gtrsim 2.9$).

For the photometric detection completeness, we used the best-fit relation from \citet{Shuntov2025a} for the F444W MAG\_MODEL magnitude in the COSMOS-Web public catalogue, 
\begin{equation}
    C(m_{F444W}) = 1 - \frac{1.07}{1+\exp{(-2.14 * ( m_{F444W} - 28.05))}} \text{\ \ \ \ ,} \label{eq:c_det}
\end{equation}
with the total completeness computed as $C(\rm{S/N}) \times C(m_{\rm{F444W}})$. We note that this is only exactly correct if the S/N of the line in the WFSS data is strictly independent of the F444W detection. Given the relative depth of the imaging and grism observations, and the fact that the F444W imaging was taken as part of COSMOS-Web, we assume this to be a reasonable working assumption.

\section{Results}
\label{sec:results}
\subsection{A sample of \oiii\ emitters at $6.75\!<\!z\!<\!9.05$ in the COSMOS field}
\begin{figure*}
    \centering
    \includegraphics[width=0.44\linewidth]{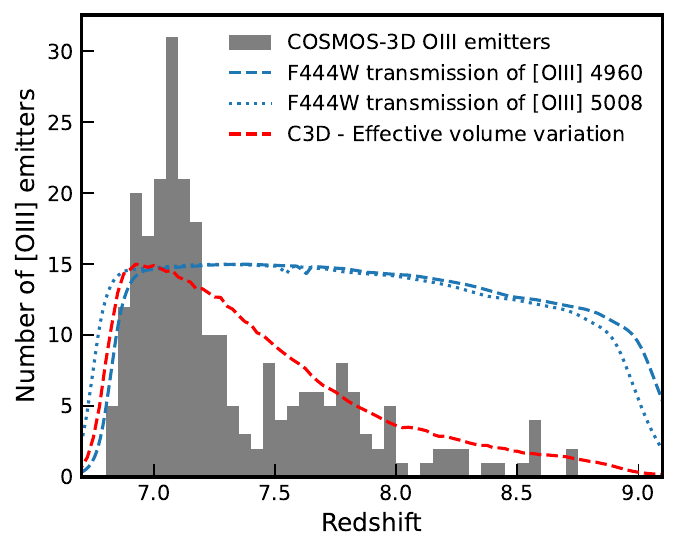}
    \includegraphics[width=0.5\linewidth]{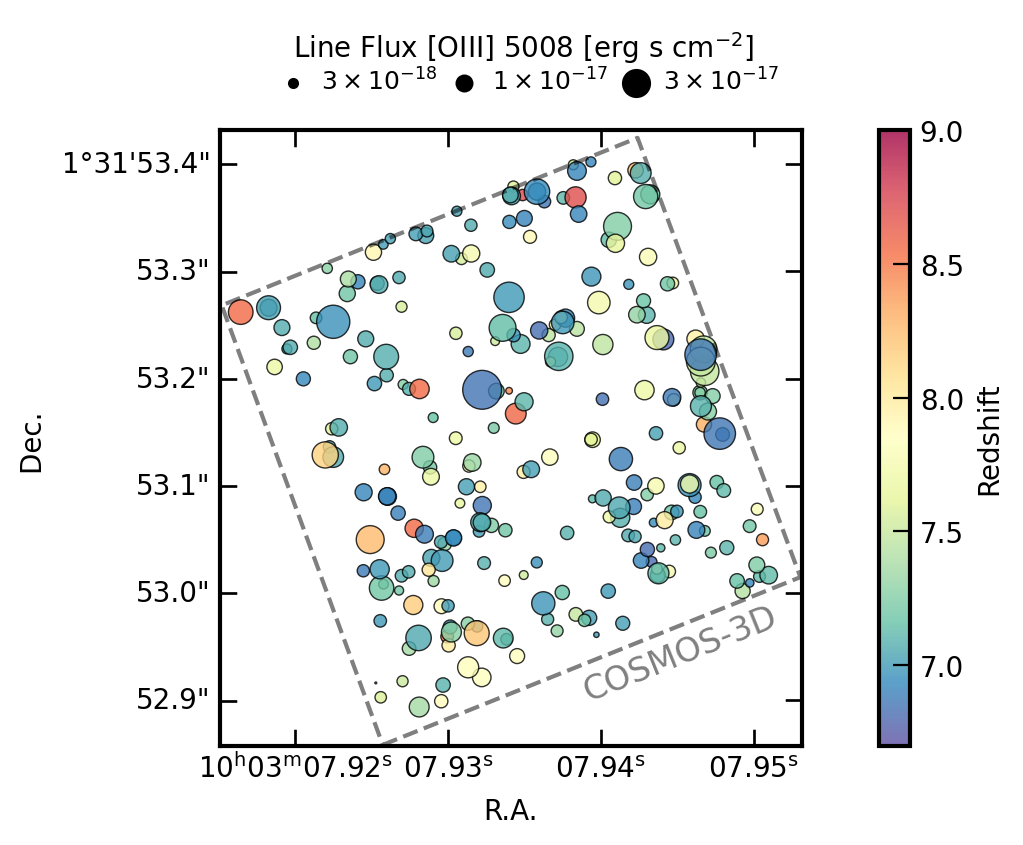}
    \includegraphics[width=0.8\linewidth,trim=0cm 5.5cm 3cm 5.8cm,clip]{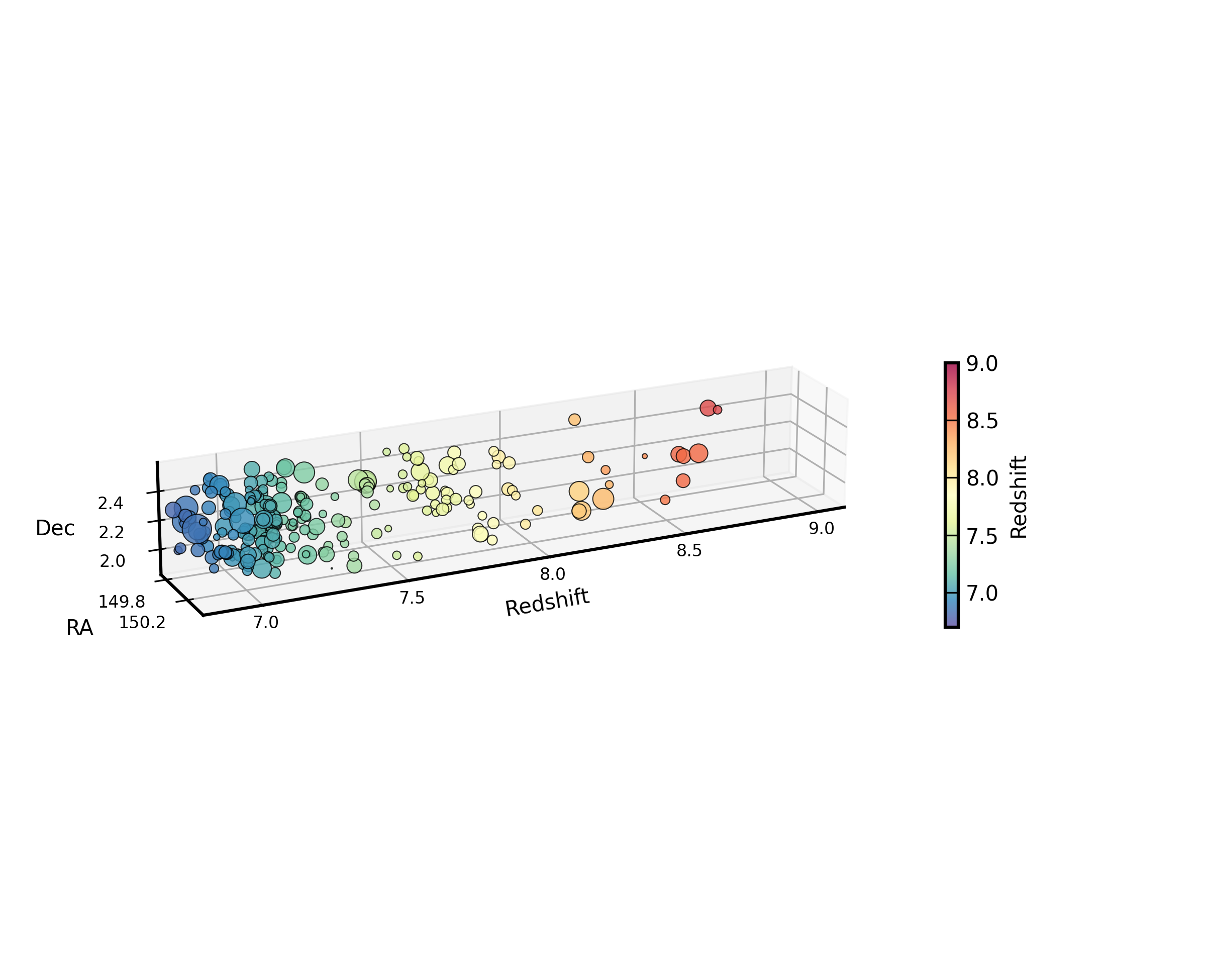}
    \caption{\textbf{Top Left:} Redshift distribution of the \oiii\  emitter sample presented in this work. We also show the coverage of the \oiii\ line in the F444W filter and the variation of the effective volume, taking into account the specific mosaic pattern of COSMOS-3D and the sensitivity of the observations (see \ref{app:rms}).\textbf{Top Right:} Sky position of the distribution of the \oiii\ emitter sample presented in this work. The sources are colour-coded as a function of redshift, and the size of the dot represents the observed \oiii\ flux. The contour of the COSMOS-3D field is outlined in grey. \textbf{Bottom: } 3D visualisation of the \oiii\ emitter sample. The size and colour scaling are the same as in the top right plot. }
    \label{fig:sample_simple_plots}
\end{figure*}

We show the redshift distribution and redshift-space of our main sample of $249$ \oiiihb\ emitters at $6.75\!<\!z\!<\!9.05$ in Fig. \ref{fig:sample_simple_plots}. The sample clearly reveal numerous structures within the COSMOS field, namely at $z\sim 7.1$ (see Champagne et al. in prep for a detailed analysis), $z\sim 7.85$, and potentially $z\sim8.2$. This is analogous to the findings of previous WFSS surveys, which all find a number of overdensities in their respective field of view \citep[e.g.][]{Wang2023, Helton2024b,Herard-Demanche2025}. Further papers will carefully study the different structures in the COSMOS field using COSMOS-3D slitless spectroscopy. We note that the present sample is particularly useful to study their spatial extent, with a $0.33\ \rm{deg}^2$ field. The full list of emitters (including line flux measurements and completeness) is available in Appendix \ref{app:full_cat} and in machine-readable format as supplementary material. We also show in Appendix \ref{app:full_cat} stacks of the \oiiihb\ emitters spectra showing no significant subtraction residuals around the \oiiihb\ and the distribution of \oiii\ 5008/4960 ratios, which we find to be in agreement with the expected value of 2.98 from atomic physics \citep{Storey2000}. 

\subsection{New constraints on the \oiii\ luminosity function}

We now proceed to compute the spectroscopically confirmed \oiii\ $5008\ \AA$ LF from our sample of \oiii\ emitters. Following \citet{Meyer2024}, we split the sample into two redshift bins: $6.75<z\leq7.50$ and $7.50<z\leq9.05$. The choice of two redshift bins and their range is motivated by 1) retaining a large number of emitters in each redshift bin, and 2) obtaining a mean and median redshift of the subsamples close to $z\sim 7$ and $z\sim8$, respectively. In this case, the two redshifts bins contain $183$ and $66$ objects at mean redshift of $\left< z \right> = 7.1,7.9$, respectively. We bin the emitters in luminosity bins of $\Delta \log_{10} L_{\oiii\ 5008} = 0.2$ dex and compute the number density using the \citet[][]{Schmidt1968} estimator in each bin,
\begin{equation}
    \Phi(L) \text{d}\log_{10} L = \Sigma_i \frac{1}{ V^{i}_{\rm{tot}} C_i} \text{\ \ \ \ ,}
\end{equation}
where $i$ is the index of each object in the luminosity bin, $V^{i}_{tot}$ is the volume in each luminosity bin. The volume where \oiii\ emitters can be detected at a given luminosity is calculated using the pseudo-rms cube constructed from all extracted spectra in the field. Importantly, this takes into account the reduced spatial coverage for $z\gtrsim 8.5$ emitters where gaps appear in the WFSS/F444W mosaic, and variations in the rms of different observations for low-luminosity sources (see Appendix \ref{app:rms}). For luminous sources ($L_{\oiii}>10^{43} \rm{erg\ s}^{-1}$), the total volume is $(6.75\!<\!z\!<\!7.5)= 1.62\times 10^6 \ \rm{cMpc}^{3},V_{tot}(7.5\!<\!z\!<\!9.05)= 2.76\times 10^6 \ \rm{cMpc}^{3}$, and $C_i$ is the completeness. Specifically, the total completeness is computed as the product of Eq. \ref{eq:c_spec} and \ref{eq:c_det} using the best-fit S/N(\oiii\ $5008$\AA) and F444W magnitude of each source.

\begin{figure}
    \centering
    \includegraphics[width=\linewidth]{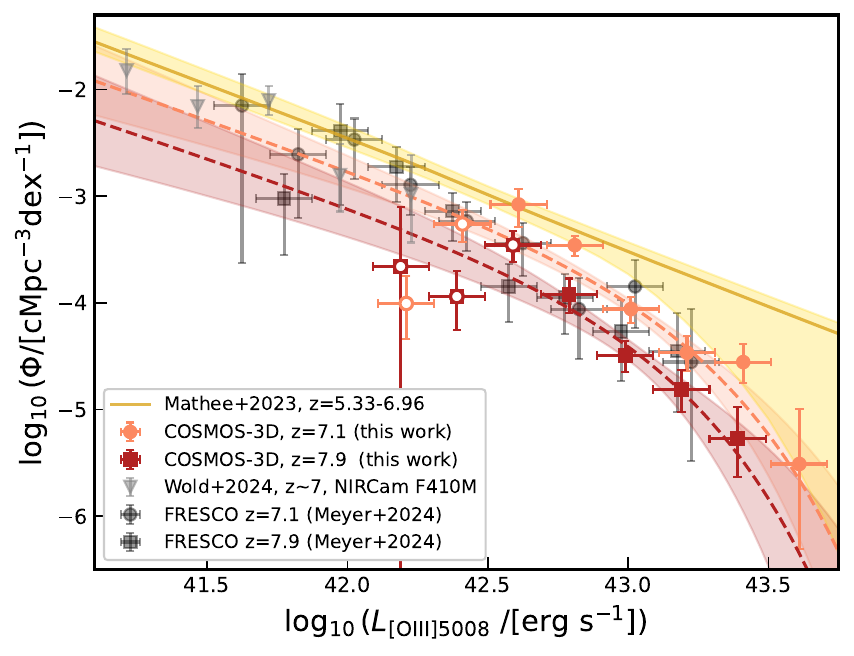}
    \caption{JWST \oiii\ luminosity functions at $5\lesssim z\lesssim 9$. The COSMOS-3D constraints determined in this work are shown with dark red and orange datapoints (empty markers indicate a completeness $<25\%$). We show the $z=5.33-6.96$ \oiii\ LF from EIGER \citep{Matthee2023_EIGER} in yellow. The datapoints at $z=7.1,7.9$ from FRESCO \citep{Meyer2024} in grey squares and circles (with corrected errors including $67\%$ cosmic variance), and the NIRCam medium band excess-selected $z\sim 7$ \oiii\ LF from \citet{Wold2025} is shown in grey triangles. The best-fit relations at $z=7.1$ and $z=7.9$ are shown in orange and red lines (and envelopes denoting the $16-84$th percentiles) which are fit to all the NIRCam WFSS observations. }
    \label{fig:O3_LF}
\end{figure}

\begin{table*}
    \centering
    \setlength{\tabcolsep}{6pt} 
\renewcommand{\arraystretch}{1.3} 
    \caption{Summary of the $z\sim7$ and $z\sim8$ [\ion{O}{iii}] LF samples and best-fit Schechter function.}
    \begin{tabular}{c|c|c|c|c|c|c}
         Redshift range & $\bar{z}$ & $\left< z\right>$ & $N_{\rm{gal}}$ & $\log_{10} \phi^{*}$ & $\log_{10} L_{\rm{[\ion{O}{iii}] 5008}}^{*}$ & $\alpha$ \\ \hline
        $6.75\!<\!z\!<\!7.50$ & 7.07 & 7.09  & 183 & $-4.10_{\tabularrel-0.49}^{\tabularrel+0.36}$ & $43.09^{\tabularrel+0.27}_{\tabularrel- 0.20}$ &
        $-1.92^{\tabularrel+0.24}_{\tabularrel- 0.24}$ \\
        $7.50\!<\!z\!<\!9.05$ & 7.78 & 7.90 & 66 & $-4.30_{\tabularrel-0.65}^{\tabularrel+0.44}$ & $42.98^{\tabularrel+0.36}_{\tabularrel-0.22}$ &  $-1.88^{\tabularrel+0.26}_{\tabularrel- 0.26}$  \\
    \end{tabular}
    \tablefoot{We give in order the redshift range, median ($\bar{z}$) and mean ($\left< z\right>$) redshift, number of emitters ($N_{\rm{gal}}$), and marginalised values of \oiii\ $5008\ \AA$ LF Schechter function parameters ($\log_{10} \phi^{*}$, $\log_{10} L_{\rm{[\ion{O}{iii}] 5008}}^{*}$, $\alpha$), for which we provide the median value and the $16-84$ percentiles of the marginalised posterior distribution (see also the full posterior distributions in Appendix \ref{app:full_O3}).}
    \label{tab:o3_lf_parameters}
\end{table*}

The variance on the number density is the sum in quadrature of the Poisson error on the number of objects in the bin, the completeness error obtained by resampling the covariance matrix of the best-fit sigmoid at the S/N of each emitter, and a relative cosmic variance error of $15\%$ (see further Section \ref{sec:cv} for our constraints on cosmic variance). 

We present the resulting \oiii\ $5008\ \AA$ LF in Figure \ref{fig:O3_LF}, alongside high-redshift JWST results  \citep{Matthee2023_EIGER, Meyer2024,Wold2025}. We find excellent agreement with the results of \citet{Meyer2024}, albeit with a slightly lower number densities, clearly indicating a smooth evolution from $z\sim 6$ \citep{Matthee2023_EIGER} to $z\sim 7,8$ (this work). Additionally, the significant increase in the field of view compared to FRESCO enables us to detect objects up to a luminosity of $\log_{10} L_{\oiii} / [\rm{erg\ s}^{-1}] \sim 43.4$, constraining the knee of the \oiii\ $5008\ \AA$ LF at $z\sim7,8$. Together with the NIRCam medium-band imaging results of \citet[][]{Wold2025,Korber2025} at the faint-end, we are able to constrain the parameters of the \oiii\ $5008\ \AA$ LF at $z\sim 7-8$.

\begin{figure*}
    \centering
    \includegraphics[width=\linewidth]{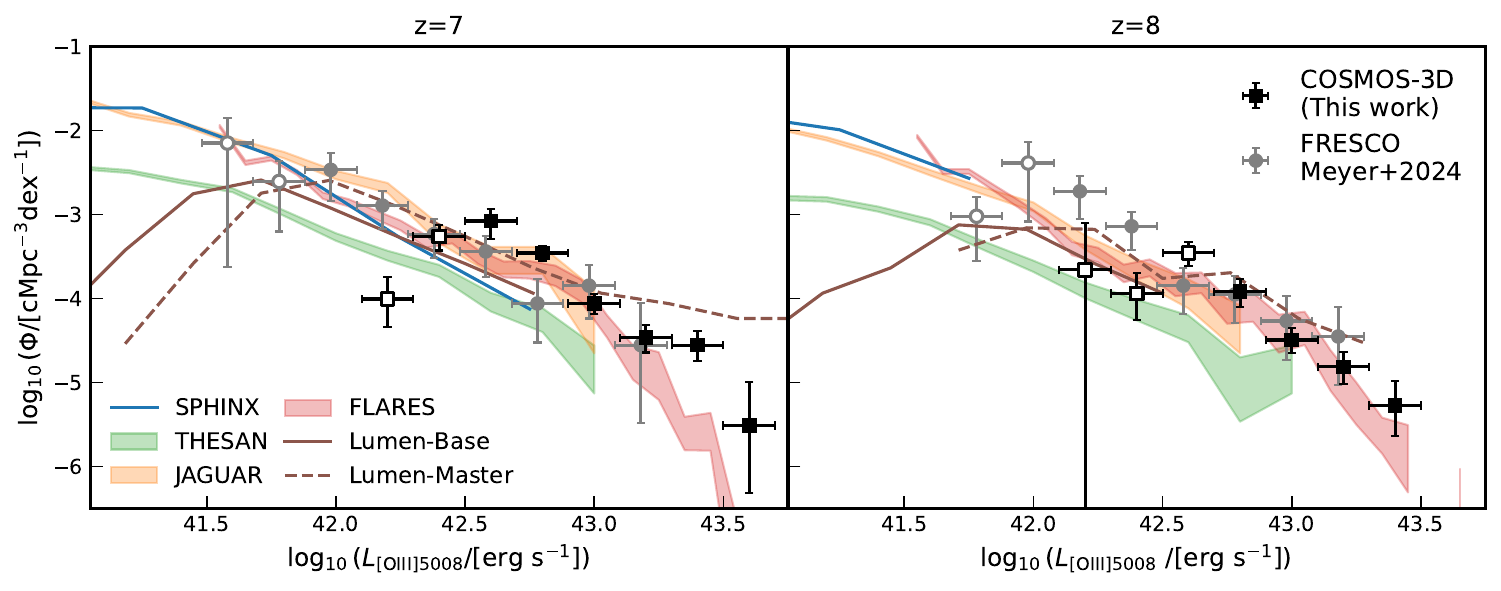}
    \caption{Comparison between the spectroscopically confirmed \oiii\ luminosity function (black and grey datapoints) and theoretical predictions at $z\simeq 7,8$. Empty datapoints indicate a low completeness ($<25\%$). The errors from \citet{Meyer2024} have been corrected for cosmic variance. The COSMOS-3D constraints are in significant tension with the THESAN and SPHIX predictions but in good agreement with JAGUAR, FLARES and Lumen(see further text). Our new observations constrain the \oiii\ LF beyond the knee ($L_{\oiii}>10^{43} \rm{erg\ s}^{-1}$) for the first time, showing good agreement with theoretical models covering this regime (FLARES, Lumen).
    }
    \label{fig:O3_LF_sims}
\end{figure*}
 
We fit simultaneously the FRESCO and COSMOS-3D \oiii\ $5008\ \AA$ LF measurements with a single Schechter function. In doing so, we correct the errors of the FRESCO results \citep{Meyer2024} by adding a $67\%$ relative error in quadrature to account for cosmic variance (see \ref{sec:cv}), and use a conservatively broad Gaussian prior $\mathcal{N}(\mu=-1.75,\sigma=0.25)$ for the faint-end slope $\alpha$ following medium-band survey results \citep[][]{Wold2025, Korber2025}. The parameters are all relatively well constrained, in contrast with earlier studies \citep{Matthee2023_EIGER,Meyer2024} reporting significant degeneracies between the parameters and the fiducial faint-slope $\alpha$ (see further Appendix \ref{app:full_O3}). We report the best-fit parameters values for the Schechter parameters $(\phi^*, L_{\oiii\ 5008}^*, \alpha$) in Table \ref{tab:o3_lf_parameters}. We find a notable improvement over the previous constraints from \citet{Meyer2024}, with the uncertainty on $\phi^*, L^*$ reduced by a factor $\sim 2-4$ at $z\sim 7-8$, respectively. We attribute all these improvements to our ability to constrain the knee of the \oiii\ $5008\ \AA$ LF with rare luminous emitters, underscoring the importance of the large Field of View of COSMOS-3D to constrain the \oiii\ $5008\ \AA$ LF. Our best-fit solution prefers a faint-end slope $\alpha(z\sim7) = -1.92^{+0.24}_{+0.24}, \alpha(z\sim8) = -1.88^{+0.26}_{-0.26}$ in between that of \citet{Korber2025} ($-1.78< \alpha < -1.55$) and that of \citet{Bouwens2015} for the UVLF at the same redshift ($\alpha=-2.06\pm 0.13,-2.03\pm0.23$) but consistent with both within uncertainties. 

\subsection{Comparison of the \oiii\ $5008\ \AA$ LF and EW distribution with theoretical predictions}

We present an updated comparison of the spectroscopically confirmed  z=7,8 \oiii\ $5008\ \AA$ LF against predictions from THESAN \citep[][]{Kannan2022a,Kannan2022}{}{}, SPHINX \citep[][]{Katz2023_sphinx}{}{}, JAGUAR \citep[][]{Williams2018}{}{}, FLARES \citep[][]{Lovell2021,Wilkins2023_o3}{}{} and for the Lumen model \citep[][which builds on the emission-line modelling from \citet{Hirschmann2023}]{Scharre2026},  in Fig. \ref{fig:O3_LF_sims}. For Lumen we show their “Base'' and “Master'' models (with dashed and full lines) which are expected to bracket the observed galaxy population \citep[see further][]{Scharre2026}. The simulations can be broadly divided into two categories with respect to their modelling of the \oiii\ line: semi-analytical models and post-processed hydrodynamical simulations (JAGUAR, FLARES,  Lumen) and radiation-hydrodynamics simulations (SPHINX, THESAN). The improvement in the observed \oiii\ $5008\ \AA$ LF compared to \citet{Meyer2024} (and the re-assessment of their cosmic variance errors) now enable us to provide more accurate constraints. Our errors are now in range of the simulations uncertainties and, for the first time, are not dominated by cosmic variance (see \ref{sec:cv}). We find in general good agreement between our observations and the results of FLARES, JAGUAR and Lumen within their dynamical range. We can now firmly point out a discrepancy between our observations and predictions from THESAN and SPHINX (although the overlap between observations and predictions is limited to a small luminosity range). We however note that whilst the bright-end \oiii\ $5008\ \AA$ LF is well reproduced by FLARES and JAGUAR, the \oiii/UV scatter and trend with luminosity is not well reproduced \citep[see e.g.][]{Meyer2024}.

\begin{figure}
    \centering
    \includegraphics[width=\linewidth]{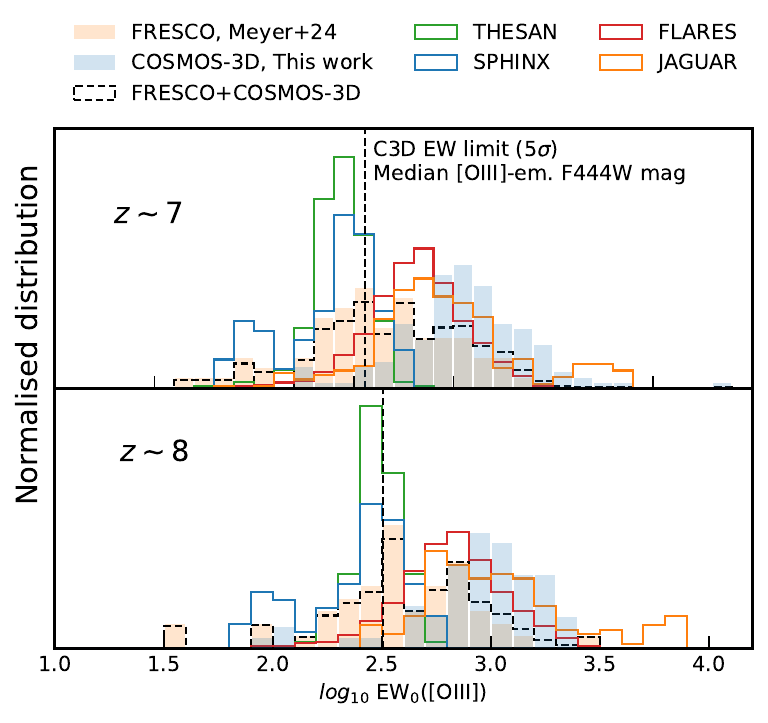}
    \caption{Equivalent width (EW$_0$) distribution of the \oiii\ emitters in this work, FRESCO \citep{Meyer2024}, and predictions from THESAN \citep[][]{Kannan2022a,Kannan2022}{}{}, SPHINX \citep[][]{Katz2023_sphinx}{}{}, JAGUAR \citep[][]{Williams2018}{}{}, and FLARES \citep[][]{Lovell2021,Wilkins2023_o3}{}{}. The combined distribution of FRESCO and COMSOS-3D, re-weighted by survey area, is shown in dashed black lines. }
    \label{fig:ew_o3}
\end{figure}

We now turn to the equivalent width (EW) distribution of $z=7,8$ \oiii\ emitters in simulations and observations. Our large field of view observations should capture the most extreme sources, making the comparison more definitive than in \citep{Meyer2024}. We measure the rest-frame equivalent width (EW$_0$) using the continuum derived from the COSMOS-Web F444W aperture magnitude, to which we subtract the line fluxes integrated over the filter width. In the following analysis, we remove the handful low-S/N objects with S/N(EW$_0)<2$. We show the observed EW distribution in Fig. \ref{fig:ew_o3} from FRESCO, COSMOS-3D and predictions from the models discussed previously (in a similar UV luminosity regime). At the median F444W magnitude of our \oiii\ emitter sample, the median COSMOS-3D grism sensitivity (across the field and wavelength range) results in a typical rest-frame equivalent width $5\sigma$ sensitivity limit of $359\ \AA$. We note that this is a estimate but in practice the EW limit is different by a factor of a few for each individual sources depending on its F444W magnitude, redshift and location in the COSMOS-3D field. The majority ($\sim 74\%$) of the sample presented is above that limit and $69\%$ of the sample have an equivalent width $\rm{EW}_0>500\ \AA$, suggesting that we are not missing a significant population of low-EW emitters at the luminosities probed here. We also combine the FRESCO and COSMOS-3D distributions by re-weighting the numbers of emitters by the relative field of view of FRESCO and COSMOS-3D (factor $\sim 9$). The resulting distribution is still incomplete at the low-EW end, especially since the depth of the FRESCO F444W imaging is shallower than the COSMOS-Web imaging data. However, we can characterise the high-EW tail where COSMOS-3D is most likely complete. 

We find that FLARES and JAGUAR clearly over-predict the number of high-EW emitters with EW$_0\gtrsim 360$\ \AA. Whilst the location of the peak and spread of the EW distribution is difficult to determine from spectroscopic surveys alone due to completeness issues, it is clear that the observed EW distribution above $>360\ \AA$ is flatter and biased towards lower values than predicted by FLARES and JAGUAR. This is consistent with the finding of \citet{Meyer2024} that the median \oiii/UV ratio is too high in FLARES and JAGUAR; assuming a flat spectrum in $f_\nu$, this translates to higher equivalent widths of \oiii. In contrast,  the THESAN and SPHINX EW(\oiii) could be closer to the observed distribution, but it is impossible to conclude with the current data. Indeed the peak and the median of the THESAN/SPHINX predicted distribution is below the EW limit of COSMOS-3D (and likely that of FRESCO), where incompleteness will lead to a precipitous decline of the number of emitters. Deeper spectroscopic observations of \oiii\ emitters (with NIRSpec or NIRCam WFSS) can however address this point.

In summary, this confirms the results of \citet{Meyer2024} where semi-analytical models or post-processed hydrodynamical simulations in large volumes produce better constraints on the global \oiii\ $5008\ \AA$ LF at the cost of over-predicting the \oiii/UV relation or the EW distribution. Interestingly, THESAN and SPHINX do not over-predict the high-end of the EW(\oiii) distribution (although uncertainties regarding the median EW which is at the limit of our observations). The combination of these different observables will therefore be of value to improve the next generation of simulations.

\section{Discussion}
\label{sec:discussion}

\subsection{Cosmic variance of $z\sim 7-9$ \oiii\ emitters}
\label{sec:cv}

\begin{figure*}
    \centering
    \includegraphics[width=0.9\linewidth]{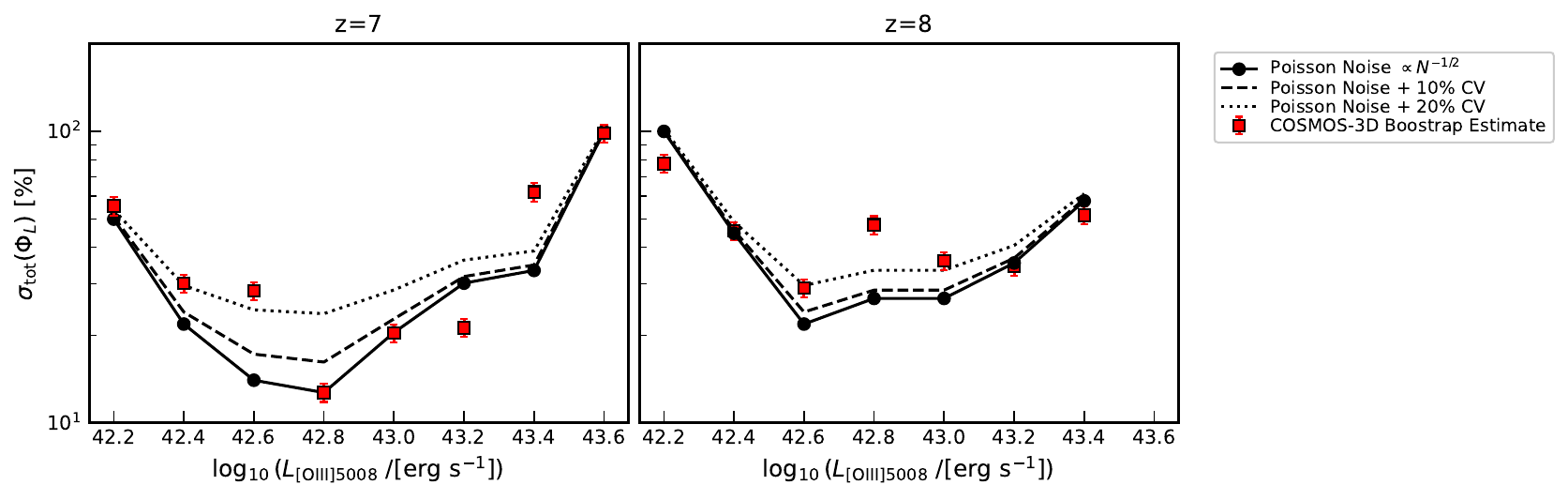}
    \caption{Bootstrap estimate of the total variance of the \oiii\ $5008\ \AA$ LF at $z\sim7$ (left) and $z\sim8$ (right). The Poisson noise variance is shown in black circles in each bin. We illustrate the expected total variance, including $10\%$ or $20\%$ cosmic variance, as dotted and dashed lines. We measure a best-fit cosmic variance (taken as the weighted difference in quadrature between the estimated variance and the Poisson variance) of $15.1\%$ at $z\sim7$ and $13.4\%$ at $z\sim8$. }
    \label{fig:bootstrap}
\end{figure*}

\begin{figure*}
    \centering
    \includegraphics[width=\linewidth]{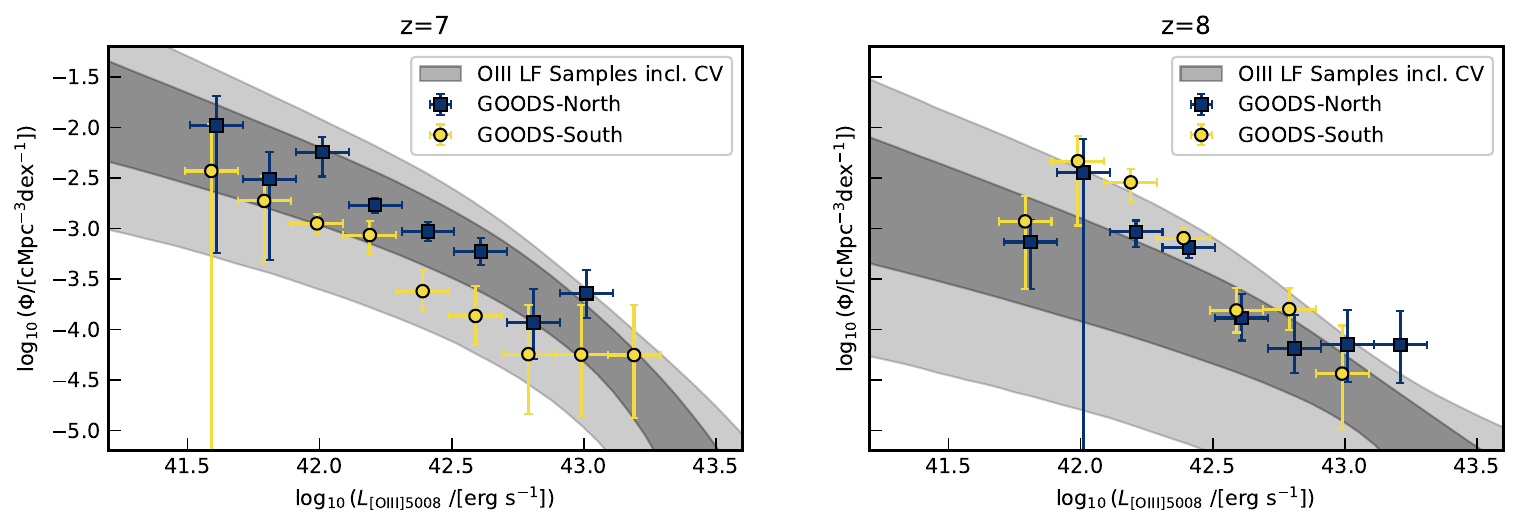}
    \caption{Comparison between the \oiii\ $5008\ \AA$ LF in the GOODS fields \citep[coloured points][]{Meyer2024} and bootstraps samples of the \oiii\ $5008\ \AA$ LF using the best-fit parameters from this work and an added cosmic variance of $47\%$ (see Section \ref{sec:cv}) in shades of grey (1-2$\sigma$ envelopes). The errors on the FRESCO measurements are presented here without the cosmic variance correction included previously. At $z=7$ GOODS-South is at the median density whilst GOODS-North is overdense at $2\sigma$. At $z=8$ we find a slight overdensity in both GOODS-South and GOODS-North, although it must be noted that constraints below $L_{\oiii\ }=10^{42.5}\ \rm{erg\ s}^{-1}$ have a low completeness and thus are more uncertain \citep{Meyer2024}. }
    \label{fig:fresco_GN_GS_context}
\end{figure*}

We now leverage the large field of view of COSMOS-3D to determine the impact of cosmic variance on the $7\!<\!z\!<\!9$ \oiii\ $5008\ \AA$ LF. We divided the COSMOS-3D data into its 20 observations that are each a mosaic of $4\times2$ NIRCam WFSS pointings (Kakiichi in prep.). Incidentally, this is the same mosaic pattern and field of view of a single FRESCO GOODS field \citep{Oesch2023}, making the comparison with earlier studies straightforward. 

We computed a bootstrap estimate of the \oiii\ $5008\ \AA$ LFs at $z=7$ and $z=8$ by using randomly sampling the $20$ observations with replacement. Assuming that the $20$ observations are independent, we find that the bootstrap estimate uncertainty is very close to the relative uncertainty due to the Poisson error uncertainty at the bright end where the sample is complete (Fig. \ref{fig:bootstrap}). Assuming that cosmic variance is not strongly luminosity-dependent over the luminosity range considered, we take the average difference (in quadrature) of the Poisson error and the estimated bootstrap uncertainty to compute a fiducial cosmic variance of $6.7\pm1.2\%$ at $z\sim7$ and $19.4\pm1.9\%$ at $z=7.9$. We note that the $z\sim 7$ estimate is impacted by the presence of a large overdensity spanning the whole COSMOS-3D field (Champagne in prep.), thus biasing our estimate lower as the bootstrap samples are (most likely) highly correlated. 

An alternative way to estimate the relative cosmic variance is to derive it directly from the galaxy bias and the dark matter cosmic \citep{Robertson2010}
\begin{equation}
    \sigma^{CV}_{\rm{[OIII]}} = b_{\rm{[OIII]}} \sigma_{DM} \text{\ \ \ \ \ ,}
\end{equation}
where we use \textit{QuickCV} \citep{NewmanDavis2002} to estimate the dark matter variance for our survey volume ($2.1,2.0\%$) for the $6.75\!<\!z\!<\!7.5, 7.5\!<\!z\!<\!9.05$ samples, respectively. We use the recent \oiii\ bias measurement from \citet{Shuntov2025_clustering} to find $\sigma^{CV}_{\rm{[OIII]}} = 16.0^{+0.5}_{-0.7}\%,15.6^{+0.5}_{-0.7}\%$ at $z\sim7,8$, respectively. We argue this is roughly consistent with our measurements if considering a few caveats. First of all, the COSMOS-3D measurement at z$=7$ is likely biased low, explaining the better agreement with our $z=8$ measurement ($19.4\pm1.9\%$). Secondly, \citet{Shuntov2025_clustering} measure the \oiii\ bias using the much smaller FRESCO fields (thus missing large angular scales) and in a smaller luminosity regime, likely resulting in a smaller bias value. Thirdly, our bootstrap samples are a biased estimator of the true cosmic variance of the \oiii\ $5008\ \AA$ LF as we use a contiguous field of view. Throughout this work we adopt a fiducial value of 15$\%$ for the cosmic variance of the \oiii\ $5008\ \AA$ LF in our two chosen redshift bins.

\subsection{Placing the overdensities in the GOODS fields in context}

\citet{Meyer2024} determined the \oiii\ $5008\ \AA$ LF in the two GOODS-South and -North fields using a similar approach as highlighted here, and found a disagreement of a factor $\sim 3$ in the total number density between the fields. It was however uncertain at the time whether one field (or both) was under-/over-dense and by how much. With a better estimate of the average \oiii\ $5008\ \AA$ LF and a cosmic variance estimate, we can now put the density of \oiii\ emitters in the two GOODS fields into context.

We scale the 15$\%$ estimate for the COSMOS-3D field by $\sqrt{20}$ for a single FRESCO field to get $67\%$ cosmic variance, and thus obtain $47\%$ for the two combined FRESCO fields. We draw samples from the best-fit parameters posterior distribution of the \oiii\ $5008\ \AA$ LF fitted uniquely to the COSMOS-3D data, and renormalise the sampled LF by a factor drawn from a normal distribution $\mathcal{N}(1,0.64)$. We then show the 1 and 2 $\sigma$ envelopes of the resulting draws against the FRESCO GOODS-North and South \oiii\ $5008\ \AA$ LF in Fig. \ref{fig:fresco_GN_GS_context}. At $6.75\!<\!z\!<\!9.05$, we find that GOODS-North is overdense at the $2\sigma$ level over the luminosity range considered,while GOODS-South is in excellent agreement with the median \oiii. At $7.50\!<\!z\!<\!9.05$ however, both GOODS-South and GOODS-North fields are slightly overdense at $\sim1\sigma$, consistent with our finding of a more rapid decline of the \oiii\ luminosity density than found in \citet{Meyer2024}. Overall, these field-to-field variations are consistent with the $67\%$ variance estimate and the observed difference between the number of emitters in each field \citep{Meyer2024}. It is thus important to note that a massive overdensity such as the one found in the FRESCO-GN field at $7.0\!<\!z\!<\!7.2$ around a faint quasar \citep{Fujimoto2022} is \textit{only} boosting the \oiii\ $5008\ \AA$ LF in the redshift range $6.75\!<\!z\!<\!7.5$ by $\sim30\%$ (a $\sim2 \sigma$ fluctuation), meaning that such overdensities should be encountered in $\sim2.5\%$ of fields surveyed. 

We conclude this section by extrapolating the relative cosmic variance for the \oiii\ $5008\ \AA$ LF at $6.75\!<\!z\!<\!7.5,7.50\!<\!z\!<\!9.05$ found in COSMOS-3D for a single NIRCam pointing ($\sim 268\%$), and a $2$x$2$ pointings mosaic ($\sim 134\%$). These large numbers illustrate the limited inferences that can be made from individual small field of view observations. Conversely, these can be viewed as an opportunity to identify even more massive overdensities than already found in existing contiguous grism surveys with a relatively small number of parallel observations distributed on the sky \citep[e.g. SAPPHIRES, POPPIES,][Kartaltepe in prep.]{Sun2025}.

\subsection{Evolution of the \oiii\ $5008\ \AA$ LF parameters over cosmic time}
\label{sec:redshift_evol}
\begin{figure}
    \centering
    \includegraphics[width=\linewidth]{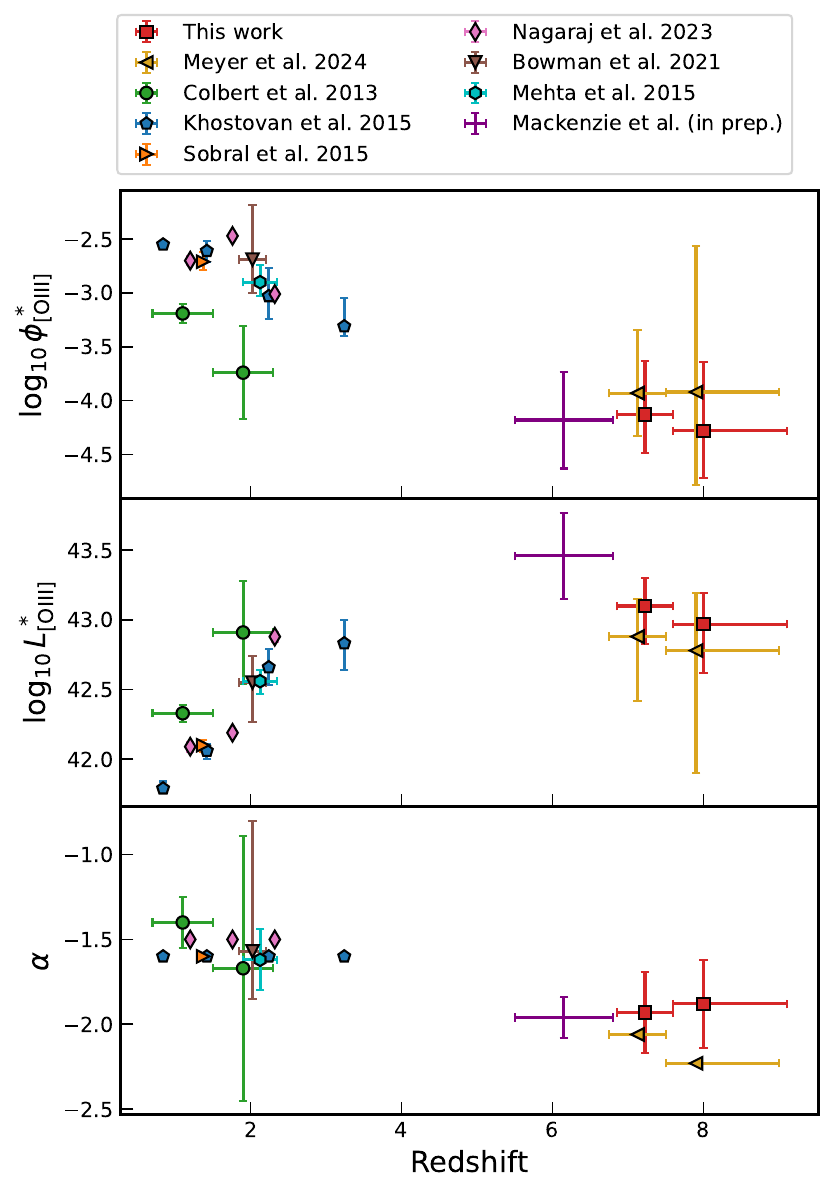}
    \caption{Redshift evolution of the \oiii\ $5008\ \AA$ LF Schechter parameters over $1\!<\!z\!<\!9$ compiled from \citet[][]{Colbert2013,Khostovan2015, Sobral2015, Mehta2015,Bowman2021,Nagaraj2023,Meyer2024}; Mackenzie et al. (in prep), and this work. We do not show the constraints from \citet{Matthee2023_EIGER} as their errors are very large in the absence of good constraints on the knee of the LF. This work shows that the evolution in the \oiii\ $5008\ \AA$ LF parameters is more rapid around Cosmic Noon $1\!<\!z\!<\!3$ than at higher redshifts, with a potential decline in the \oiii\ luminosity density at z$\gtrsim 7$ needing further confirmation (see text). }
    \label{fig:o3_LF_evolution_redshift}
\end{figure}

The improvement in the constraints on the \oiii\ $5008\ \AA$ LF Schechter parameters enables us to compare our results to studies of \oiii\ emitters at cosmic noon and later epochs. We show a compilation of the normalisation, typical luminosity and faint-end slope of the \oiii\ $5008\ \AA$ LF in Fig \ref{fig:o3_LF_evolution_redshift}. Our results show that the \oiii\ $5008\ \AA$ LF normalisation and decrease smoothly between $z\sim 2$ and $z\sim 6$ and then remains constant at $6<z<8$. By contrast $L^{*}_{[OIII] 5008}$ seems to reach a peak around $z\sim 6$ and then decline at higher redshift. Whilst the constraints on the faint-end slope are more scarce, there is no evidence for a significant evolution, in stark contrast with that found for the UV LF for the same redshift range \citep[e.g.][]{Bouwens2015}. The absence of evidence for an evolution of the \oiii\ $5008\ \AA$ LF faint-end slope is in agreement with results from lensed low-luminosity \oiii\ emitters detected in GLIMPSE \citep{Korber2025}. We caution against over-interpretation since the three parameters are still quite degenerate (see Appendix \ref{app:full_O3}) and better constraints on the faint-end are needed to make further progress. Unfortunately, cosmic variance will dominate in the small fields used for deep, faint-end measurements, making the combination of the bright-end and faint-end results difficult. 

We finally discuss the integrated \oiii\ luminosity density, which is better constrained than the individual parameters of the Schechter function. Integrating the \oiii\ best-fit LF parameter from \citet{Matthee2023_EIGER} results in a \oiii\ luminosity density of $\rho_{[\ion{O}{iii}]} (z\sim6)=$ $(1.1^{+0.30}_{-0.29})\times10^{40}\ \rm{erg\ s\ }^{-1}\rm{cMpc}^{-3}$. Using the same integration limit, we find \oiii\ luminosity densities of $\rho_{[\ion{O}{iii}]} (z\sim7)=$ $(1.78^{+0.30}_{-0.29})\times10^{39}\ \rm{erg\ s\ }^{-1}\rm{cMpc}^{-3}$ and $\rho_{[\ion{O}{iii}]} (z\sim 8)=$ $(0.80^{+0.27}_{-0.21}) \times10^{39}\ \rm{erg\ s\ }^{-1}\rm{cMpc}^{-3}$. We thus have clear evidence for a decline (factor $\times 10$) between $z\sim6$ and $z\sim 8$, and tentative evidence ( $2.5\sigma$) for $\times 2.2$ decline in the \oiii\ luminosity density between $z\simeq7$ and $z\simeq8$ (e.g. in only $100$ Myr). This is expected if the weak evolution of the \oiii\ luminosity density between $z\sim2$ \citep{Colbert2013, Khostovan2015} and $z=7$ is due to a combination of changes in metallicity or temperature compensating the decline in star-formation rate density (SFRD) \citep[e.g.][]{Matthee2023_EIGER,Meyer2024}. However, at $z\sim 6-9$ \oiii\ emitters found JWST slitless surveys have \oiii/H$\beta$ ratios close to the maximum of $\sim 7$ \citep[e.g.][]{Bian2015,Nakajima2022}. Consequently, any decrease in the metallicity beyond $z>8$ would only add to the decrease of the SFRD, thus resulting in a very rapid decline of the \oiii\ luminosity density at z$\gtrsim8$. If a decline in the \oiii\ luminosity density was confirmed, thus indicating a divergence from the UV LF at z$>8$, this could be used to chart the decline of chemical enrichment at early times. Wide-area (e.g. $>0.1\ \rm{deg}^2$) MIRI imaging surveys with a wide wavelength coverage would probably be more suited to chart this decline than deeper MIRI imaging surveys with much smaller field of view \citep[e.g.][]{Rieke2024, Ostlin2025} given the importance of cosmic variance uncertainties highlighted in this work.

\section{Conclusions}
\label{sec:conclusion}
We have presented a census of \oiiihb\ emitters in the COSMOS field over the $\sim 0.33 \ \rm{deg}^2$ field covered by NIRCam/WFSS F444W slitless spectroscopy the JWST COSMOS-3D programme (\#5893). Our results are as follows:
\begin{itemize}
      \item We report the discovery of 249 \oiii\ emitters at $6.75\!<\!z\!<\!9.05$ with \oiii\ luminosity $42 <\log_{10}L_{\rm{[OIII]} 5008} / [\rm{erg\ s}^{-1}\rm{cm}^{-2}] \lesssim 43.5$. Simultaneously, we have determined the completeness of our sample, ideal for follow-up studies.
      \item Rare, extremely luminous \oiii\ emitters are captured thanks to the extended field of view of COSMOS-3D, thus constraining the bright end of the \oiii\ luminosity function. In conjunction with complementary constraints on the faint end from medium-band imaging survey, we can determine the knee, normalisation and slope of the \oiii\ $5008\ \AA$ LF at $z=7-8$ with improved accuracy. In particular, we find evidence for a decline in the \oiii\ luminosity density between $z\sim6$ and $z\sim8$, which we attribute to the decreasing metallicity of \oiii\ emitters at $z\gtrsim7$.
      \item The improved accuracy of our \oiii\ $5008\ \AA$ LF measurement and the constraints at the bright-end are in good agreement with predictions from semi-analytical models and post-processed hydrodynamical simulations (JAGUAR, FLARES,  Lumen). In contrast, radiation-hydrodynamics simulations (SPHINX, THESAN) underpredict the \oiii\ $5008\ \AA$ LF in the luminosity regime probed ($L_{\oiii} \gtrsim 10^{42}\ \rm{erg\ s})^{-1}$. However, the latter simulations match the EW distribution more closely than the former, which predict large EW emitters beyond EW(\oiii)$\gtrsim 3000\ \AA$ which are not found in COSMOS-3D despite a larger field of view and thus better coverage of rare, extreme objects.
      \item We leverage our wide area survey to measure the cosmic variance of the \oiii\ $5008\ \AA$ LF at $6.75\!<\!z\!<\!9$. We estimate the cosmic variance to be $\sim 15\%$ for a volume of $1.62(2.76)\times 10^{6}\ \rm{cMpc}^3$ at $z\sim 7(8)\ (\Delta z = 0.75,1.55)$. In most luminosity bins, our errors are thus not dominated by cosmic variance. In contrast, the uncertainties on the \oiii\ $5008\ \AA$ LF in smaller field of view surveys (e.g. FRESCO, CONGRESS and EIGER) is dominated by cosmic variance. We further show that the differences in the number of \oiii\, emitters between the GOODS-South and -North FRESCO fields are completely consistent with the $\sim 67\%$ cosmic variance expected for their smaller field of view. 
\end{itemize}

\section{Data availability}

Table \ref{tab:full_cat} is available in electronic form at the CDS via anonymous ftp to cdsarc.u-strasbg.fr (130.79.128.5) or via \url{http://cdsweb.u-strasbg.fr/cgi-bin/qcat?J/A+A/}.

\begin{acknowledgements}
The authors thank the anonymous referee for constructive comments and suggestion which improved this work. RAM thanks R. Mackenzie for sharing the parameters of the upcoming \oiii\ $5008\ \AA$ LF based on EIGER, and for general discussions on fitting the \oiii\ $5008\ \AA$ LF. RAM acknowledges support from the Swiss National Science Foundation (SNSF) through project grant 200020\_207349. FW, BJ acknowledge support from NSF award AST-2513040. K.K. acknowledges support from VILLUM FONDEN (71574). The Cosmic Dawn Center is funded by the Danish National Research Foundation under grant no. 140. JBC  acknowledges funding from the JWST Arizona/Steward Postdoc in Early galaxies and Reionization (JASPER) Scholar contract at the University of Arizona.
IFAE is partially funded by the CERCA program of the Generalitat de Catalunya. The research of SS was supported by the European Union’s Horizon Europe research and innovation programme (COSMO-LYA, grant agreement 101044612). YX acknowledges support from JSPS KAKENHI Grant Number JP25KJ1029. YL acknowledges support from JSPS KAKENHI Grant Numbers 24K17084 and 25H00663.  This work has received funding from the Swiss State Secretariat for Education, Research and Innovation (SERI) under contract number MB22.00072.
\\
This work is based on observations made with the NASA/ESA/CSA James Webb Space Telescope. The data were obtained from the Mikulski Archive for Space Telescopes at the Space Telescope Science Institute, which is operated by the Association of Universities for Research in Astronomy, Inc., under NASA contract NAS 5-03127 for JWST.  These observations are associated with programs \#5893. Support for program \#5893 was provided by NASA through a grant from the Space Telescope Science Institute, which is operated by the Association of Universities for Research in Astronomy, Inc., under NASA contract NAS 5-03127. We acknowledge the strong support provided by the program coordinator Alison Vick and instrument reviewers Brian Brooks and Jonathan Aguilar. Support for this work was provided by NASA through grant JWST-GO-01727 awarded by the Space Telescope Science Institute, which is operated by the Association of Universities for Research in Astronomy, Inc., under NASA contract NAS 5-26555.
\end{acknowledgements}

\bibliographystyle{aa}
\bibliography{library_C3D_O3LF}

\onecolumn
\appendix

\section{Sensitivity limits and wavelength coverage}
\label{app:rms}
\begin{figure*}
    \centering
    \includegraphics[width=0.8\linewidth]{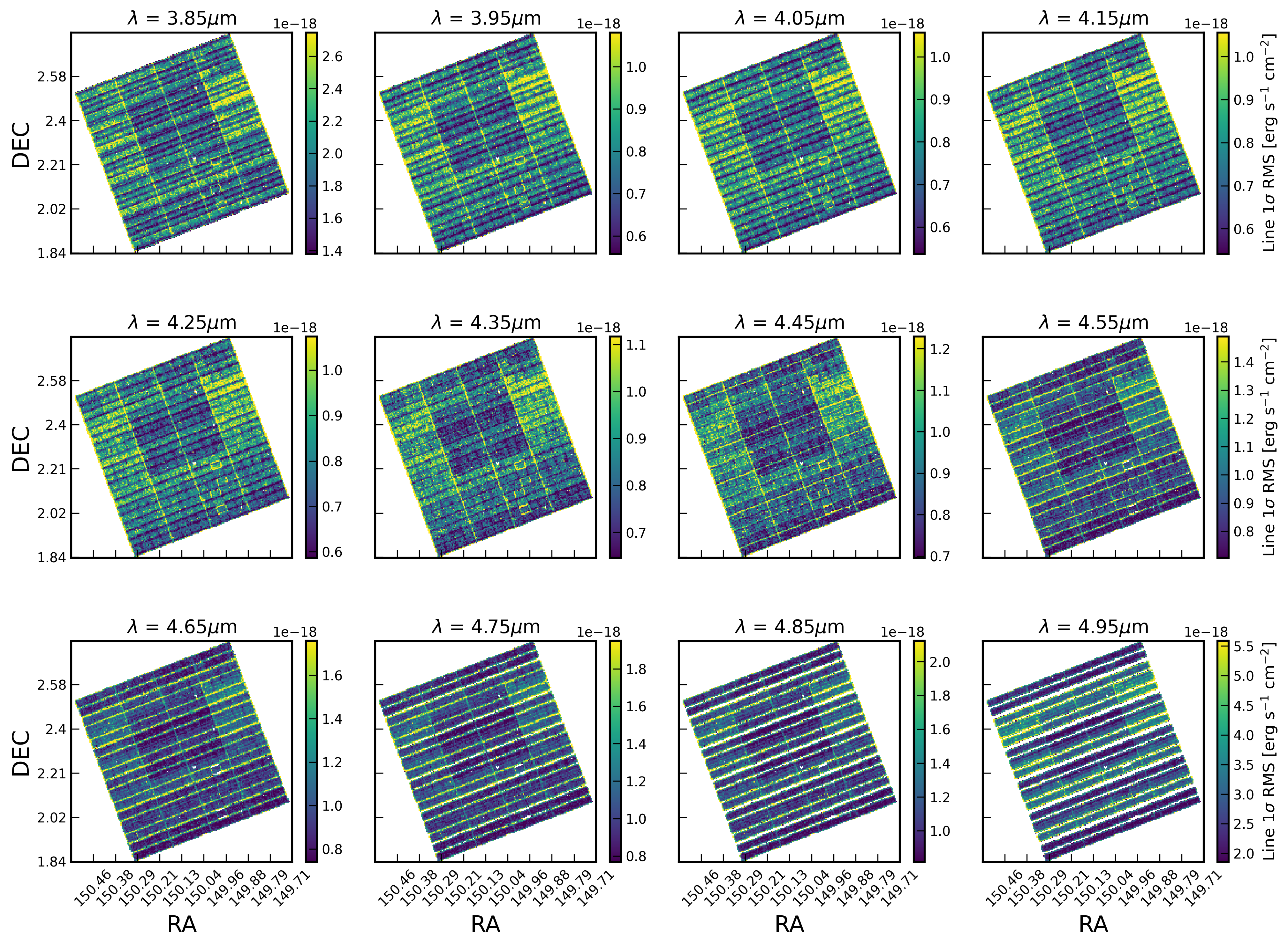}
    \caption{Line sensitivity limits across the field and for different wavelengths, integrated over FWHM=$100\ \rm{km\ s}^{-1}$ line width broadened by the wavelength-dependent grism resolution.}
    \label{fig:rms}
\end{figure*}

We show in Fig. \ref{fig:rms} the line sensitivity limit (assuming an intrinsic FWHM=$100\ \rm{km\ s}^{-1}$ broadened by the LSF) across the COSMOS-3D footprint (as of Summer 2025). The sensitivity is calculated using the error array from the optimally extracted spectra of all COSMOS-Web galaxies covered by COSMOS-3D, rescaled to match the rms noise measured in the same spectrum. We then take the median rms at a given RA, Dec, $\lambda$ to create a pseudo-rms cube. Importantly, gaps appear at $\lambda>4.8\mu\rm{m}$, meaning that the spatial coverage of \oiii\ emitters decreases at $z\gtrsim 8.5$.

\section{Full catalogue of \oiii\ emitters}
\label{app:full_cat}
\begin{figure}
    \centering
    \includegraphics[width=0.48\linewidth]{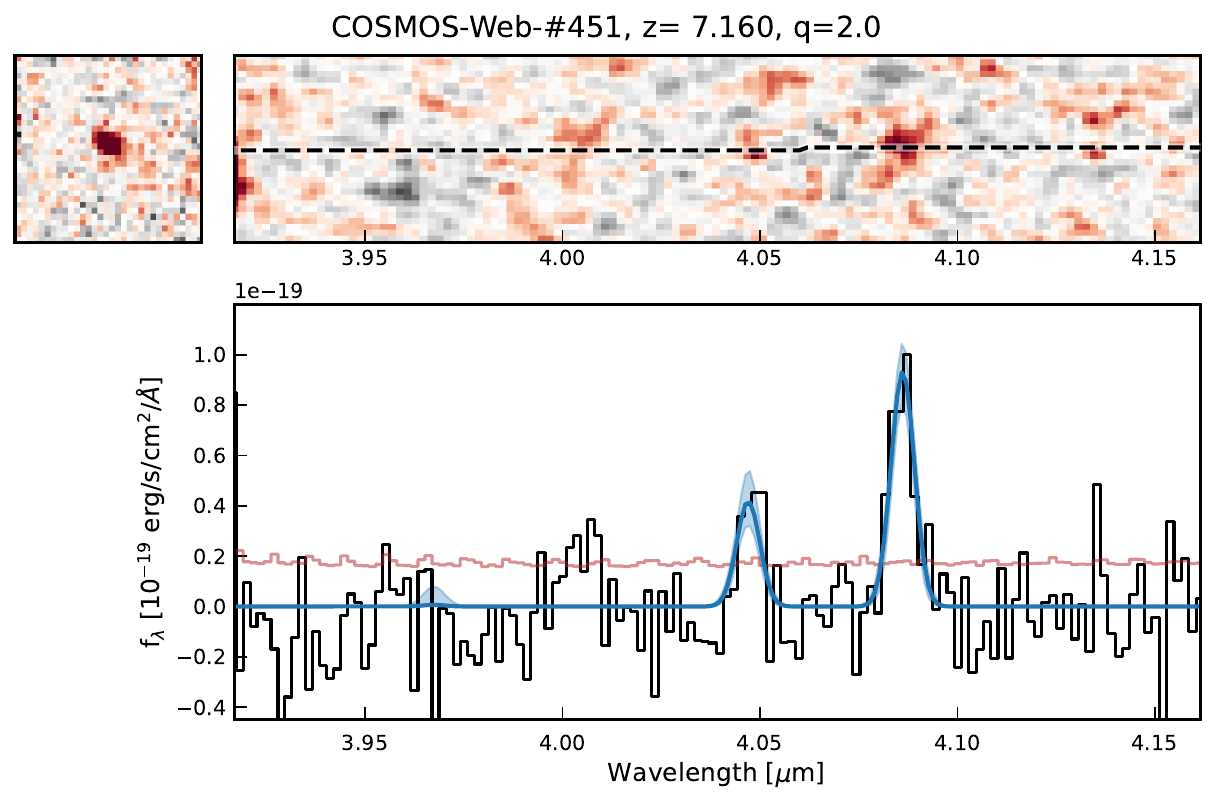}
\includegraphics[width=0.48\linewidth]{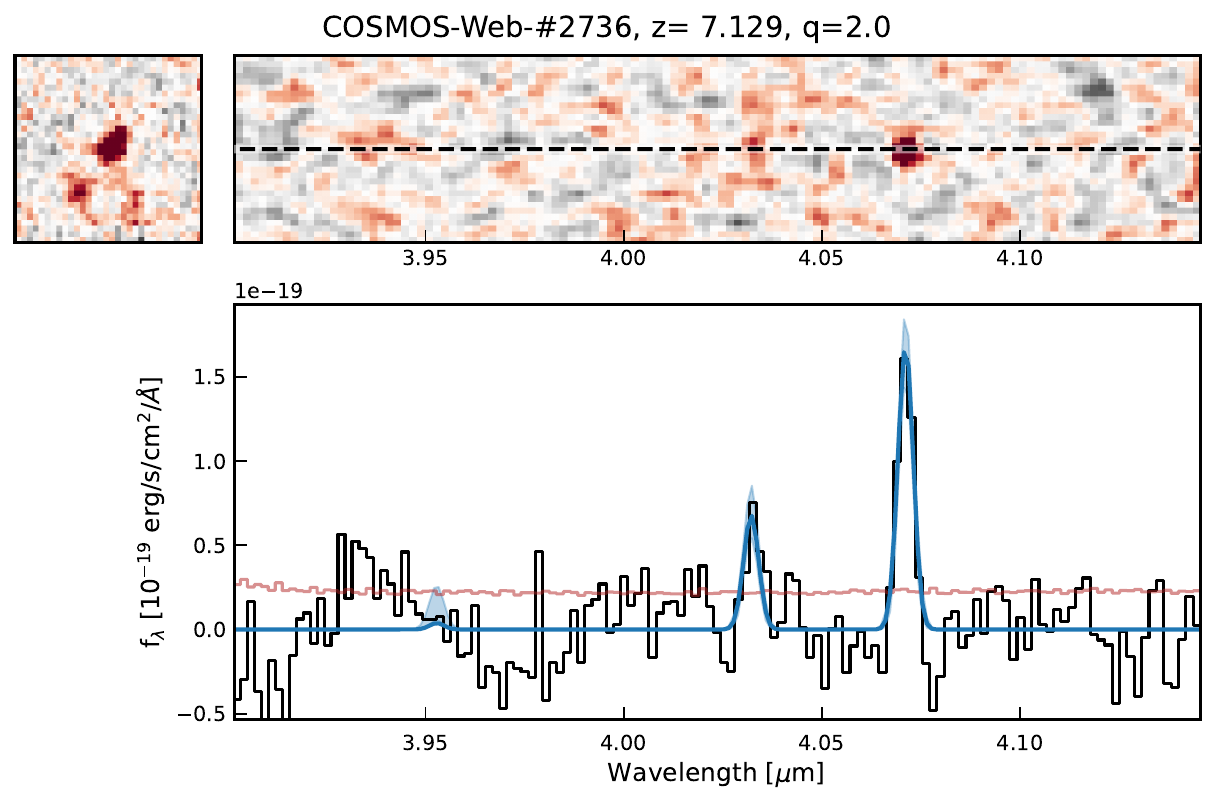} \\
\includegraphics[width=0.48\linewidth]{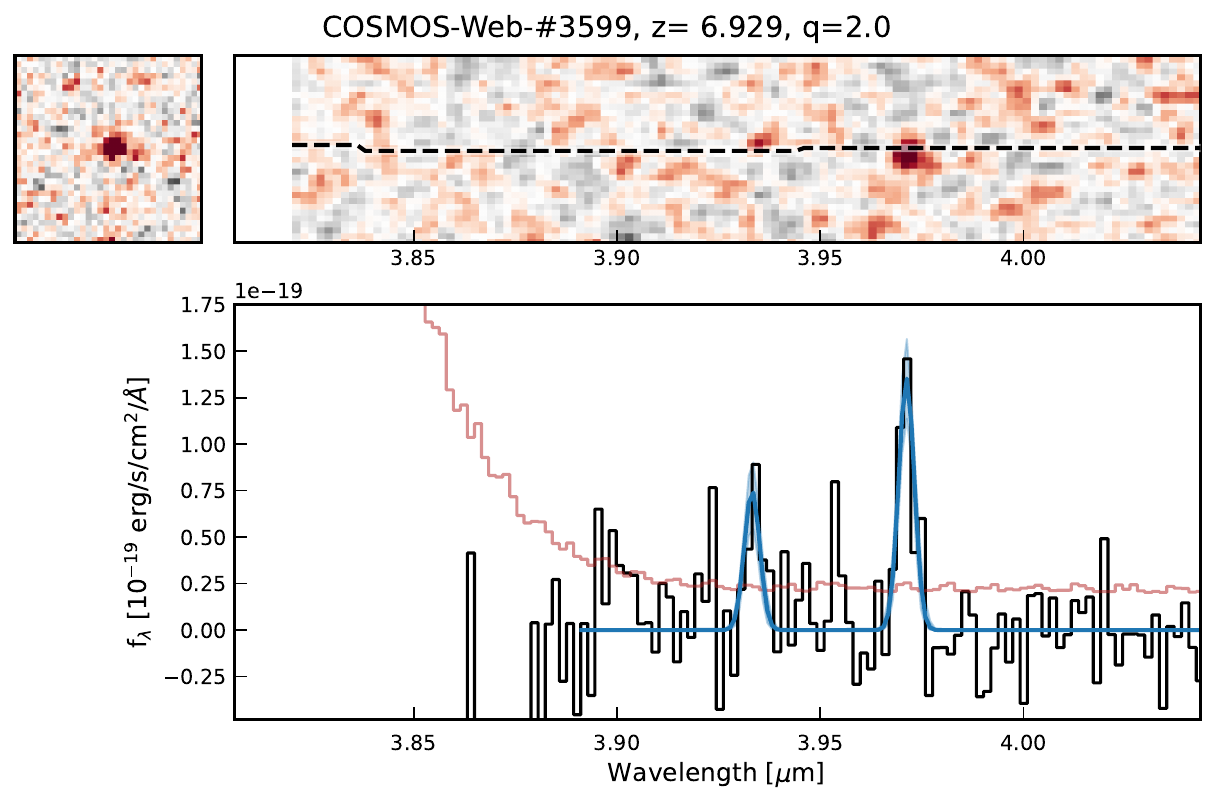}
        \includegraphics[width=0.48\linewidth]{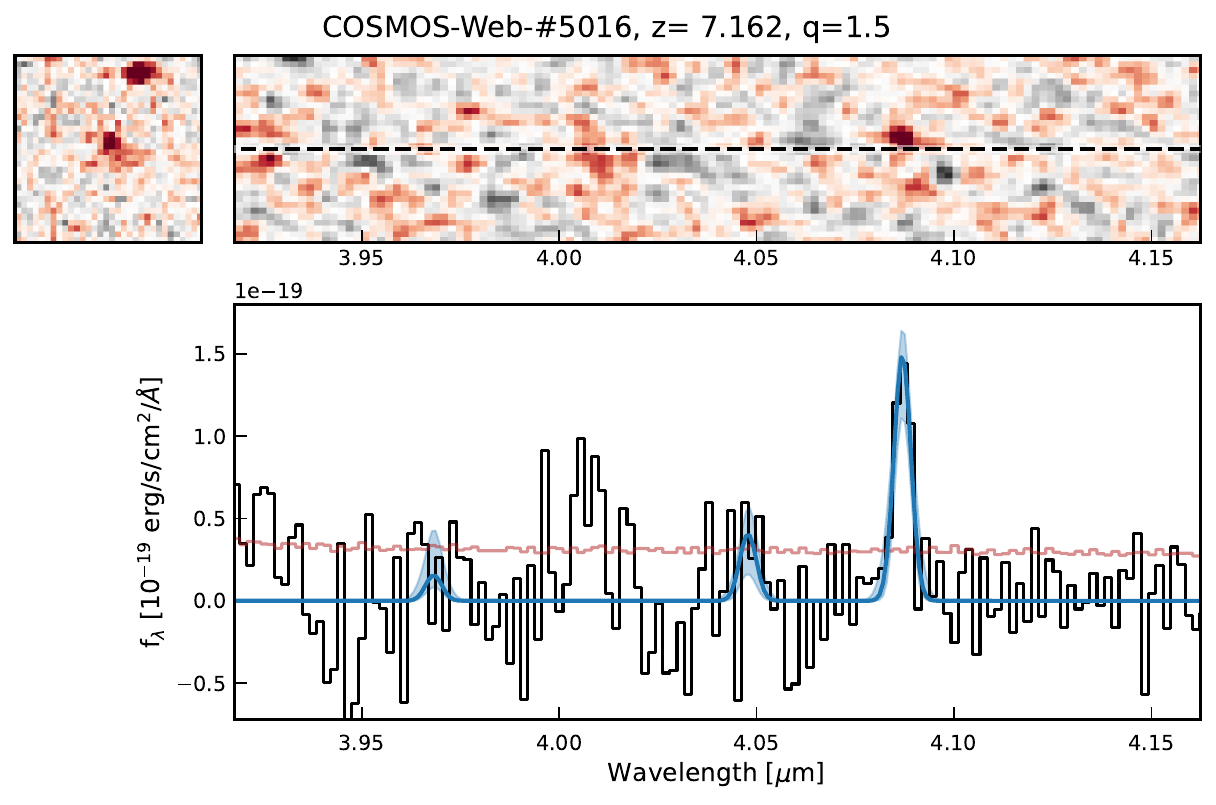} \\
    \caption{Direct F444W image stamp (top left), 2D (top right) and optimally extracted 1D (bottom panel) spectra of the four first emitters in the catalogue. }
    \label{fig:o3_appendix_selected}
\end{figure}

In this appendix we present the full catalogue of \oiii\ emitters used in this study. All presented emitters are listed using their COSMOS-Web IDs \citep{Shuntov2025_DR}, as well as their fluxes and completeness in Table \ref{tab:full_cat}. We show plots of the direct F444W imaging, 1D and 2D spectra for a short selection of objects in Fig \ref{fig:o3_appendix_selected}. Table \ref{tab:full_cat} is available in electronic form at the CDS via anonymous ftp to cdsarc.u-strasbg.fr (130.79.128.5) or via \url{http://cdsweb.u-strasbg.fr/cgi-bin/qcat?J/A+A/}, and also at \url{https://github.com/rameyer/cosmos3d}.

\begin{table*}
   
    \centering
    \tiny
   \caption{Abridged catalogue of \oiii\ emitters presented in this work. }
   \setlength{\tabcolsep}{6pt} 
   \renewcommand{\arraystretch}{1.3} 
    \begin{tabular}{c|c|c|c|c|c|c|c|c}
    ID & RA & Dec & z & $\bar{q}$ & $f_{\rm{H\beta}}$ & $f_{\rm{[OIII] 4960}}$ & $f_{\rm{[OIII] 5008}}$ & Completeness  \\ 
  C-WEB & [deg] & [deg] & &  & [$10^{-18}$ erg s$^{-1}$ cm$^{-2}$] & [$10^{-18}$ erg s$^{-1}$ cm$^{-2}$] & [$10^{-18}$ erg $s^{-1}$ $cm^{-2}$] &  \\ \hline 
$451$ & $149.840551$ & $2.114942$ & $7.159$ & $2.0$ & $0.05\pm0.26$ & $3.10\pm0.74$ & $6.93\pm0.89$ & $0.34\pm0.07$ \\ 
$2736$ & $149.809834$ & $2.141376$ & $7.129$ & $2.0$ & $0.18\pm0.51$ & $3.28\pm0.79$ & $8.16\pm0.94$ & $0.46\pm0.09$ \\ 
$3599$ & $149.847329$ & $2.133055$ & $6.929$ & $2.0$ & -- & $3.61\pm0.90$ & $6.51\pm0.96$ & $0.21\pm0.05$ \\ 
$5016$ & $149.818947$&	$2.151639$ & $7.160$  & $1.5$& $0.83 \pm 0.11$ & $2.17\pm1.23$	& $8.00\pm1.52$ & $0.10\pm0.04$ \\
... & & ... & & ... &  & ... &  & ... \\
\end{tabular}
\tablefoot{We report, in order, the COSMOS-Web ID, the RA and Dec coordinates, the redshift, the median visual inspection quality flag $q$, the \oiiihb\ fluxes and errors, and the end-to-end completeness and error.}
 \label{tab:full_cat}
\end{table*}

\begin{figure*}
    \centering
    \includegraphics[width=0.8\linewidth]{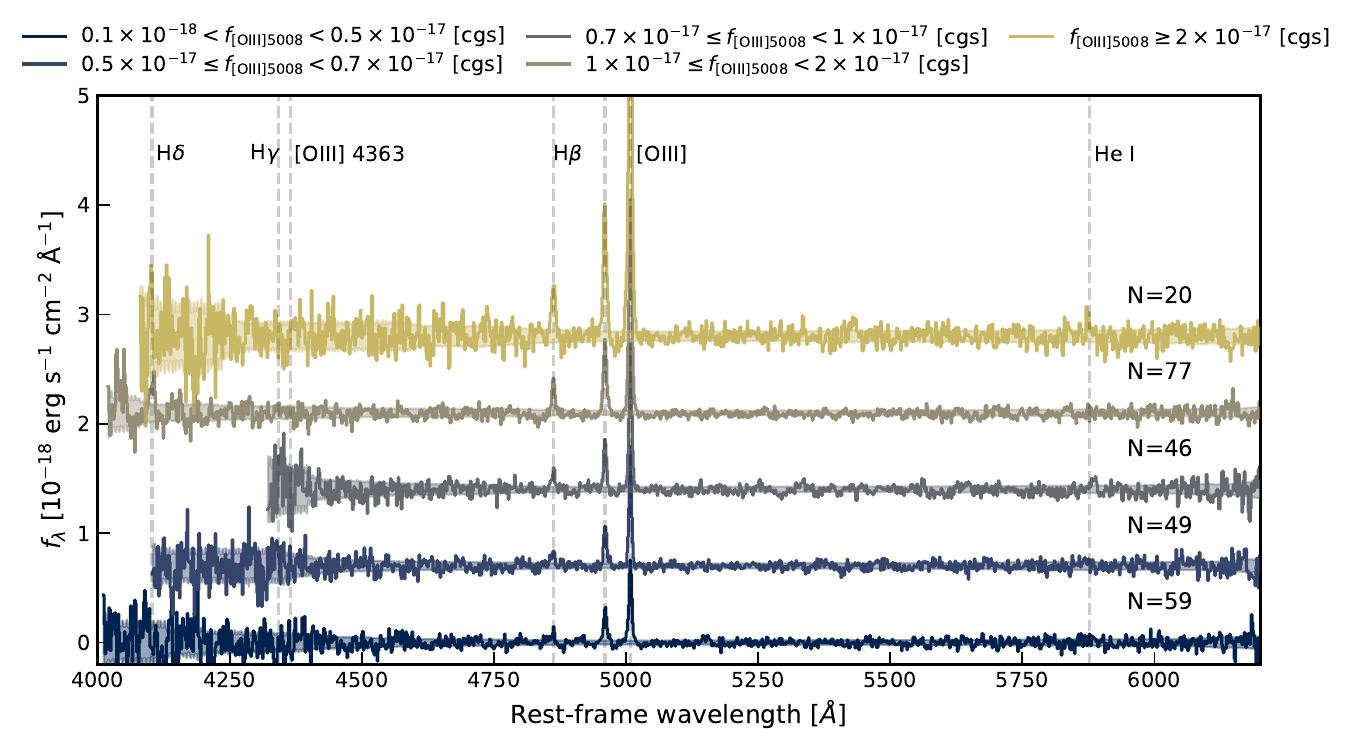}
    \caption{Stacks of the NIRCam/WFSS spectra of the \oiii\ emitters presented in this work. We find no evidence for over-subtraction around the \oiii\ line, and detect H$\beta$ even in the faintest stack, evidencing the high purity of our sample. }
    \label{fig:stacks}
\end{figure*}

\begin{figure}
    \centering
    \includegraphics[width=\linewidth]{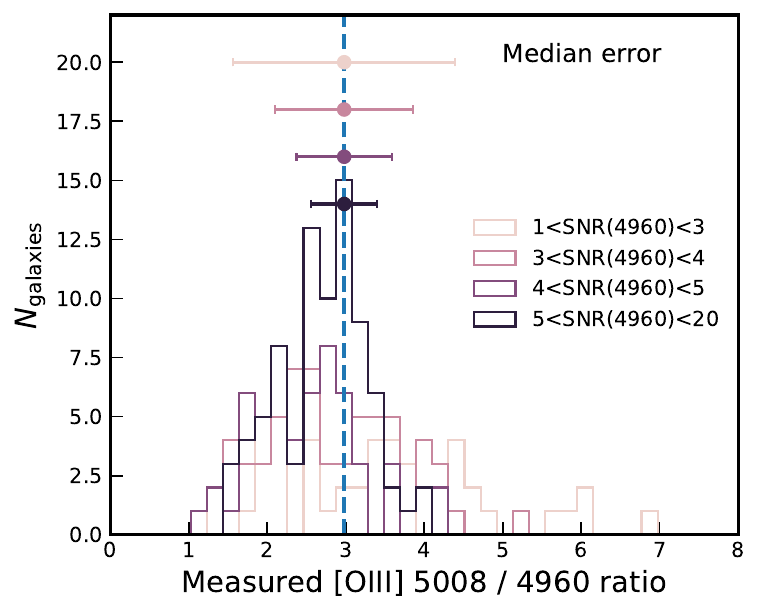}
    \caption{\oiii\ 5008/4960 ratio for the sample of emitters presented in this work, as a function of \oiii\ 4960 S/N. The scatter around the expected atomic physics value (2.98) is consistent with the median measurement error for each subsample (coloured errorbars). }
    \label{fig:o3_ratio}
\end{figure}

We also show stacks of the \oiii\ profiles as a function of the \oiii\ 5008 line flux in Fig. \ref{fig:stacks}. The stacks are produced using median stacking and inverse-weigthing by the luminosity distance. We find no significant negative wings or wiggles around the main \oiiihb\ lines, demonstrating the effectiveness of our improved continuum subtraction. We further show the distribution of measured \oiii\ 5008/4960 ratios in Fig. \ref{fig:o3_ratio}. The observed scatter is consistent with the observational uncertainties with a median ratio close to the expected value of 2.98.

\twocolumn

\section{Full \oiii\ $5008\ \AA$ LF measurements and Schechter Posterior distributions}
\begin{figure}
    \centering
    \includegraphics[width=\linewidth]{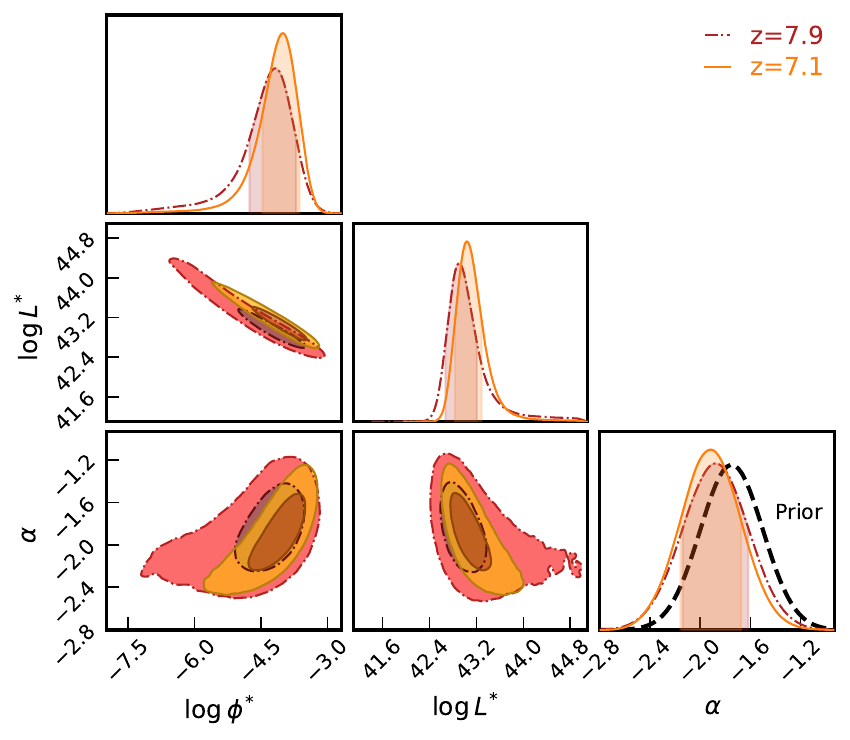} 
    \caption{Full posterior distribution of the inferred Schechter parameters of the \oiii\ $5008\ \AA$ LF at $z=7,8$ from this work. We show the prior on the faint-end slope ($\mathcal{N}(-1.75,0.25)$) in dashed black. }
    \label{fig:o3_posterior}
\end{figure}

In this appendix we give the binned values of the \oiii\ $5008\ \AA$ LF at $z=7.1,7.9$ in Table \ref{tab:o3_lf_values} and show the full posterior distribution of the \oiii\ Schechter function parameters in Fig. \ref{fig:o3_posterior}.

\begin{table}
    \centering
        \setlength{\tabcolsep}{6pt}
\renewcommand{\arraystretch}{1.3} 
    \caption{Tabulated values of the \oiii\ $5008\ \AA$ LF.}
    \begin{tabular}{c| c|c|c|c|c} 
  &$\log_{10} L_{\rm{[OIII]}5008}$ & $N$ & $<c>$ & $log\Phi_{corr}$ \\ \hline 
z=7.1 & 42.2 &  4 & 0.13 & $ -4.01^{+0.26}_{-0.33} $ \\ 
&42.4 &  21 & 0.21 & $ -3.26^{+0.13}_{-0.17} $ \\ 
&42.6 &  51 & 0.37 & $ -3.08^{+0.14}_{-0.21} $ \\ 
&42.8 &  62 & 0.67 & $ -3.46^{+0.08}_{-0.10} $ \\ 
&43.0 &  24 & 0.83 & $ -4.06^{+0.11}_{-0.13} $ \\ 
&43.2 &  11 & 0.93 & $ -4.47^{+0.15}_{-0.18} $ \\ 
&43.4 &  9 & 0.94 & $ -4.56^{+0.17}_{-0.19} $ \\ 
&43.6 &  1 & 0.94 & $ -5.51^{+0.52}_{-0.80} $ \\ \hline
z=7.9 &42.2 &  1 & 0.01 & $ <2.54\ (2\sigma)$ \\  
&42.4 &  5 & 0.10 & $ -3.94^{+0.24}_{-0.31} $ \\ 
&42.6 &  21 & 0.18 & $ -3.46^{+0.13}_{-0.16} $ \\ 
&42.8 &  14 & 0.35 & $ -3.92^{+0.15}_{-0.18} $ \\ 
&43.0 &  14 & 0.75 & $ -4.49^{+0.14}_{-0.16} $ \\ 
&43.2 &  8 & 0.87 & $ -4.81^{+0.18}_{-0.21} $ \\ 
&43.4 &  3 & 0.93 & $ -5.27^{+0.30}_{-0.36} $ \\ 
    \end{tabular}
    \tablefoot{We give the number of \oiii\ emitters, average completeness and completeness-corrected \oiii\ density for in \oiii\ luminosity bins of $\Delta L_{\oiii}=0.2$ at z=$7.1$,$7.9$. Errors include cosmic variance and limits are at the $2\sigma$ level.}
    \label{tab:o3_lf_values}
\end{table}

\label{app:full_O3}

\end{document}